\documentclass[manuscript]{aastex}

\usepackage{booktabs}
\usepackage{rotating}

\newcommand{\hale}{ (C/1995 O1) Hale-Bopp}

\newcommand{\cg}{67P/Churyumov-Gerasimenko}

\newcommand{\hyak}{(C/1996 B2) Hyakutake}

\newcommand{\cam}{a$^3\mathrm{\Pi}$}
\newcommand{\camt}{a$^3\mathrm{\Pi}$ $\rightarrow$ A$^1\mathrm{\Pi}$}

\newcommand{\cod}{CO$_2$}
\newcommand{\wat}{H$_2$O}
\newcommand{\wati}{H$_2$O$^+$}

\newcommand{\ols}{O($^1$S)}

\newcommand{\old}{O($^1$D)}
\newcommand{\cld}{C($^1$D)}
\newcommand{\olg}{O($^3$P)}
\newcommand{\clg}{C($^3$P)}
\newcommand{\grat}{G/R ratio}

\usepackage{hyperref}
\hypersetup{
unicode=false,          
pdftoolbar=true,        
pdfmenubar=true,        
pdffitwindow=false,     
pdfstartview={FitH},    
pdftitle={},    
pdfauthor={Susarla Raghuram},   
pdfsubject={},   
pdfcreator={TeXlive},   
pdfproducer={}, 
pdfkeywords={},     
pdfnewwindow=true,      
colorlinks=true,        
linkcolor=black,         
citecolor=black,        
filecolor=magenta,      
}

\shorttitle{UV and Visible emissions on comet 67P}
\shortauthors{Raghuram et al.}

\begin{document}


\title{Prediction of forbidden ultraviolet and visible emissions  in comet \cg}


\author{Susarla Raghuram \altaffilmark{1}, Anil Bhardwaj\altaffilmark{2},
 and Marina Galand\altaffilmark{1}}
\affil{\altaffilmark{1}Department of Physics,
        Imperial College London, Prince Consort Road,
        \\London SW7 2AZ, United Kingdom.}
\affil{$^2$Space Physics Laboratory, Vikram Sarabhai Space Centre, 
    Trivandrum, 695022, India.}

%
%
%


\altaffiltext{1}{Corresponding author : raghuramsusarla@gmail.com}


\begin{abstract}
Remote observation of spectroscopic emissions is a potential
tool for the identification and quantification of various species in
comets. CO Cameron band (to trace \cod)  and atomic oxygen emissions (to
trace \wat\ and/or CO$_2$, CO) have been used to probe neutral composition
 in the cometary coma.  Using a 
 coupled-chemistry emission model, various
excitation processes
controlling CO Cameron band and different atomic oxygen and atomic carbon 
 emissions have been modelled in comet \cg\ {at 
 1.29~AU (perihelion)  and at 3~AU heliocentric distances,} which is being explored by 
  ESA's Rosetta mission.  
The intensities of CO Cameron band, atomic oxygen and atomic carbon emission 
lines as a function of projected distance are calculated
 for different CO and \cod\ volume mixing ratios relative to water.  
Contributions of different excitation processes controlling these emissions 
are quantified.
We assess how \cod\ and/or CO volume mixing ratios with respect to \wat\ can be 
derived based on the observed intensities of CO Cameron band,
atomic oxygen, and atomic carbon emission lines. 
{The results presented in this work
serve as base line calculations to understand the behaviour of low out-gassing 
cometary coma and compare them with the higher gas production rate cases (e.g. comet
Halley). Quantitative analysis of different excitation processes governing the spectroscopic
 emissions is essential 
to study the chemistry of inner coma and to derive neutral gas composition.} 
\end{abstract}




\section{Introduction}
\label{intro}
Exploration of comets with space missions is critical to probe the coma 
and to access the detailed features of a cometary nucleus. However, 
remote spectroscopic observations can provide ample information about the global 
composition
of comets. The coma composition associated with main species
can be constrained from the analysis of  airglow emissions 
using ground- and space-based telescopes. This nevertheless requires the quantitative 
assessment 
of physical processes that govern these emissions in the coma.  

Among many observed cometary ultraviolet and visible airglow spectra, metastable emission lines 
have gained special interest. Solar resonance fluorescence is not an effective excitation 
mechanism to populate excited metastable states in the coma 
 due to optically forbidden transitions. The dissociative excitation by photons, 
 suprathermal electrons such as photoelectrons, and 
thermal recombination of ions are the main channels for producing various species in 
metastable excited states. By observing the emissions from the daughter products, which
are particularly from metastable state,  the estimation of mixing ratios of  their 
respective parent species has been done in several comets \citep{Feldman04,Bockelee04}.
The lifetime of a metastable excited species is shorter ($\sim$0.7 s for \ols, $\sim$110 s 
for \old,  and $\sim$3 ms for CO(\cam)) compared to their respective parent species 
(e.g., for \wat\ it is $\sim$8~$\times$~10$^4$~s$^{-1}$ at 1 AU), and 
cannot travel  large 
radial distances in the  coma from the place of formation without being lost through the 
emission of photons or quenched in collision.
Hence,  these forbidden emissions are good tracers to quantify the gas production rates of
respective parent species  in the coma. 

Though water is the dominant species in comets, 
it is very difficult to access  cometary \wat\ infrared emissions from ground-based 
observatories because of strong absorption by terrestrial water molecules. However,
the spatial profiles of cometary \wat\ have been observed by ground-based telescopes by 
observing non-resonance fluorescence emissions \citep{Mumma95,Mumma96,Russo00}. 
Since \wat\ does not have any transitions in ultraviolet and visible regions, the forbidden 
emissions of its dissociative metastable products ([OI] 6300, 6364, 5577, 2972~\AA) have 
been used as tracers.  
Oxygen atoms that are produced in the $^1$S state,  decay 95\% of the time to 
the $^1$D state emitting photons at 5577~\AA\ (green line) wavelength, 
 while 5\% of them decay directly to the ground $^3$P state which yields 
2972 and 2958~\AA\ emission lines. The radiative decay of  $^1$D $\rightarrow ^3$P
leads to  6300 and 6364~\AA\ emission lines (red-doublet emission). 
Thus, if the green line is present in a cometary spectrum, the 
red-doublet must also be present, although the red-doublet can be formed without the green 
line.
The quantification of  \wat\ in the cometary coma has been done 
by observing [OI] 6300~\AA\ emission in many comets
\citep{Delsemme76,Delsemme79,Fink84,Schultz92,Morgenthaler01}. The direct 
de-excitation of \ols\ yields 2972~\AA\ emission line which has been observed
in cometary spectrum only once \citep{Festou81}.
Since there are other oxygen bearing species, such as \cod\ and CO, which can also 
produce these metastable states via dissociative excitation reactions,  the observed green 
to  red-doublet emission intensity ratio (hereafter referred as \grat) has been used to 
confirm the parent source of these
emission lines \citep{Cochran84,Cochran08,Cochran01,Morrison97,
Zhang01,Furusho06,Capria05,Capria08,Capria10,Decock13,McKay12,McKay13,
McKay15}.
Based on \ols\ and \old\ photorates calculated by \cite{Festou81},
the observed high \grat\ values ($>$ 0.1) were ascribed to large \cod\ and CO volume 
mixing ratios in the coma \citep{Furusho06,Capria10,McKay12,Decock13}.

The  spin forbidden atomic oxygen emission line [OI] 1356~\AA\ has also been 
detected using rocket- and space-borne UV spectrometers in comets 
\citep{Woods86,Sahnow93,Mcphate99}. 
Since  fluorescence efficiency (g-factor) for this transition is small to 
explain the observed intensity, production sources for this emission are attributed to 
electron 
impact excitation mechanisms \citep{Cravens78}. \cite{Bhardwaj96} accounted for 
various electron impact excitation sources to explain the observed emission intensity 
in comet Halley, which had gas production rate of 1.3~$\times$~10$^{30}$~s$^{-1}$ 
at 1 AU.

 \cod\ is also an important oxygen bearing species in the cometary coma but
is difficult to detect directly in the visible and ultraviolet cometary spectra because of the lack 
of electronic transitions. In order to quantify \cod\ in comets, CO Cameron band emissions 
(\camt) have 
been used as a tracer of \cod\ by assuming that the excited metastable state \cam\ \citep[lifetime is
$\sim$3 ms,][]{Gilijamse07} originate primarily 
from 
photodissociation of \cod\ 
\citep{Weaver94,Weaver97,Feldman97}. This spin-forbidden electron transition 
(\camt)
yields a band emission in the ultraviolet spectral range 1800--2600~\AA. 
Using Hubble Space Telescope (HST), 
\cite{Weaver94} detected this band emission on comet
103P/Hartley 2. The observation of CO Cameron band
emission on comet 103P led to the re-examination of  International Ultraviolet Explorer
(IUE) observed spectra and \cite{Feldman97} identified this emission  on four 
comets (viz.,  C/1979 Y1 (Bradfield), 1P/Halley, C/1989 X1 (Austin), C/1990 K1 (Levy)).
The volume mixing ratios of CO relative to water have been derived in these comets by assuming that  
photodissociation of \cod\ is the main source of CO(\cam).
However, besides photodissociation of \cod,  CO(\cam) can also be produced via electron 
impact and dissociative recombination of  CO-bearing species 
\citep{Weaver94,Bhardwaj11,Raghuram12}. 

Atomic carbon [CI] 1931~\AA\ emission line has been observed in several comets 
\citep{Feldman76,Smith80,Feldman80,Feldman97,Tozzi98}. The excited state 
of this emission is a metastable state C($^1$D) which has a lifetime of about 4080 s
\citep{Hibbert93}. The radiative decay of carbon from $^1$D to the ground $^3$P state results in 
photons at 9823 and 9850~\AA, which are analogous to atomic oxygen red-doublet emissions
at 6300 and 6464~\AA. The  [CI] 9850~\AA\ emission line has been detected in comet
Hale-Bopp by \cite{Oliversen02}. The carbon atom in $^1$D metastable state fluoresces the solar
photons at 1931~\AA\ before radiative decay to the ground $^3$P state can occur. Most of the emission line 
intensity is attributed to photodissociative excitation of CO in comets 
\citep{Feldman78, Bhardwaj99a}. 
Hence this emission line is a good tracer for CO production rate
in comets \citep{Oliversen02}. The model developed by \cite{Bhardwaj99a} calculated
 this emission line intensity in comet Halley, which is  smaller by a 
  factor 5 than that observed by IUE, and suggested  an
 involvement of additional carbon-bearing species in the coma.

In order to derive the parent species production rates in the coma based on the observed 
forbidden emission
intensities  a quantitative study of various processes that govern these emissions is 
necessary. 
We have  developed a coupled chemistry-emission model  which 
accounts for major production and loss reactions 
 of \olg, \ols, \old, \clg, \cld, and CO(\cam) in  cometary comae  
\citep{Bhardwaj96,Bhardwaj99a,Bhardwaj11,Bhardwaj12}.
 This model has been applied on several
  comets and results have been compared  with the Earth-based observations
\citep{Bhardwaj96,Bhardwaj99a,Bhardwaj11,Bhardwaj12,Raghuram12,Raghuram13,
Raghuram14, Decock15}. The
model calculations for comets 103P/Hartley~2 and 1P/Halley  have shown
that suprathermal electron impact is an important excitation process in the
formation of CO(\cam), which is more important than photodissociation of \cod\
\citep{Bhardwaj11,Raghuram12}. The model applied to study atomic oxygen emission 
lines  in comets has shown that the collisional quenching in the inner coma can
significantly change the observed \grat\
\citep{Bhardwaj12,Raghuram13,Raghuram14,Decock15}. The model calculations in active 
comets, such as \hyak\ and \hale, have shown that the \grat\ varies as a function of projected
distance and depends on the collisions in the cometary comae
\citep{Bhardwaj12,Raghuram13}. Recently, we have applied our model for the analysis of
 high-resolution spectroscopic observations made from ESO very large telescope (VLT) on 
 four comets (viz., C/2002 T7 (LINEAR),
73P-C/Schwassmann-Wachmann 3, 8P/Tuttle, and 103P/Hartley 2). This study has 
allowed to constrain the \cod\ volume mixing ratios in these comets \citep{Decock15}.

After a successful rendez-vous in August 2014,  ESA's Rosetta spacecraft is exploring 
 comet \cg\
(here-after, referred as 67P)   by escorting it from $\sim$4~AU towards  perihelion 
at 1.29~AU reached in the summer 2015.
Assessing the chemical evolution of the cometary coma, as the  comet  approaches the 
Sun, is one of the main aims of Rosetta mission. In support to Rosetta,  different 
space- and ground-based
observation campaigns are taking place to understand the spatial distribution of
different volatile species in the coma. We apply our coupled~chemistry-emission model to
comet 67P to identify and quantify the processes driving formation and loss of 
 CO Cameron band, atomic oxygen forbidden emissions and [CI] 1931~\AA\ emission line
 at perihelion and at 3~AU heliocentric distance. 
Model calculations are necessary 
 to understand the physical processes governing these metastable emissions 
and to  derive \cod\ and CO mixing ratios. The motivation for this work is to provide 
a theoretical support for the interpretation of  Earth-based  and Rosetta-based 
UV observations of comet 67P.

We present the model input parameters, which may represent the gaseous environment of 
comet 67P at its perihelion and at 3~AU, in section~\ref{model-inputs}. The modelled  various
production and destruction profiles for  different electronic states of atomic 
oxygen, and atomic carbon, and CO(\cam), and emission intensities as a function of 
projected distance are presented in section~\ref{results}. The implications of modelled 
emission profiles for comet 67P are discussed in section~\ref{discussion}. Conclusions are 
given in section~\ref{conclusions}.

\section{Model input parameters} 
\label{model-inputs}
The coupled~chemistry-emission model has been
used in the present study  which accounts for main production and loss processes for 
CO(\cam),
\olg, \old, \ols,  \clg, and \cld\  species in the inner cometary coma, as described in the 
earlier  works \citep{Bhardwaj11,Bhardwaj12,Bhardwaj90,Bhardwaj96, 
Bhardwaj99a,Bhardwaj03}. Calculations are done at  perihelion distance (1.29~AU).
The H$_2$O out-gassing rate at perihelion is assumed to be 
1~$\times$~10$^{28}$~s$^{-1}$ \citep{Snodgrass13}. The number density relative to 
water  is taken to be 5\%
for both CO$_2$ (hearafter $\mu_w$(\cod)) and  CO (hearafter $\mu_w$(CO)) as a standard neutral composition.
The on-board OSIRIS instrument on Rosetta mission observed 67P's nucleus has
bi-lobed structure with dimension 
2.5~$\times$~2.5~$\times$~2.0~km for the small lobe and 4.1~$\times$~3.2~$\times$~1.3~km for the large lobe
\citep{Sierks15,Lee15}. 
In our model we have assumed a spherical nucleus of 2~km radius for simplification. 
The neutral atmosphere is calculated  using Haser's formula \citep{Haser57} which 
assumes spherical expansion of coma with a constant  velocity of 1~km~s$^{-1}$. 
The electron temperature profile, which is required to calculate electron-ion 
recombination rate, is assumed to be same as on Halley 
\citep{Korosmezey87}.

We vary \cod\ and CO mixing ratios  to quantify the change in
 contributions of different productions and loss processes yielding  CO(\cam), \ols, \old, 
 \cld,  and
 atomic carbon and atomic oxygen in ground 
 states. The incident solar flux is  based on the measurements
from the Thermosphere Ionosphere Mesosphere Energetics and Dynamics
(TIMED)/Solar EUV Experiment (SEE) \citep{Woods05} on 2 January 2005 (for solar
activity phase with F10.7 = 100~$\times$~10$^{22}$~Wm$^{-2}$~Hz$^{-1}$) at Earth and 
scaled to heliocentric distance of 1.29~AU. It is expected to be representative
of conditions encountered in summer 2015 near perihelion from solar decreasing active 
period \citep{Vigren13}. The theoretical water collisional zone of comet 67P, with gas production rate
10$^{28}$~s$^{-1}$, is around 2000~km \citep{Whipple76}. The calculations
presented in this work are relevant for the inner coma and inside the diamagnetic
cavity. 

{We also made calculations at heliocentric distance of 3~AU assuming
 the total gas production rate of 5 $\times$ 10$^{25}$ s$^{-1}$ \citep{Gulkis15, Hassig15, Bieler15}.
 The neutral coma composition is assumed to be 80\% \wat, 15\% CO and 5\% \cod. 
 The solar flux on 1 November 2014 is used
 in the model and scaled to 3~AU using inverse square of heliocentric distance.
 The electron temperature-dependent reactions play a minor role in governing the
 intensities of these emission lines.}
 
Recently ROSINA/DFMS has made several important discoveries such as the D/H ratio \citep{Altwegg15},
 presence of N$_2$ \citep{Rubin15} and O$_2$  \citep{Bieler15} in 67P's coma. The 
 observation of molecular oxygen has an important implication in the production of \ols\ and \old.
 \cite{Bieler15} found that the local abundance of molecular oxygen is varying between 1 and 10\% around the
 67P nucleus relative to \wat\ production rate. The mean value of  molecular oxygen abundance in 67P coma is
  3.8 $\pm$ 0.85\% relative to H$_2$O production rate \citep{Bieler15}.
  In order to quantify the contribution of 
 molecular oxygen on the forbidden emission lines we 
 have taken 4\% molecular oxygen relative to \wat\ production rate in the model.
 Hence we also have done a case study by assuming the water production rate 
 5 $\times$ 10$^{25}$ s$^{-1}$ and  25\% CO, 8.3\% CO$_2$, and 4\% O$_2$ as relative abundances with respect to 
 \wat\ for the month of August 2014. These abundances are  mostly in agreement  with the ROSINA measurements between August and October 2014 \citep{Hassig15,Leroy15,Bieler15}  when the
 comet was between 3 and 3.5  AU from the Sun.

\section{Results} 
\label{results}
\subsection{Formation and destruction  of CO(\cam)}
\label{results_cam}
\begin{figure}[h]
   \centering
   \includegraphics[scale=1]{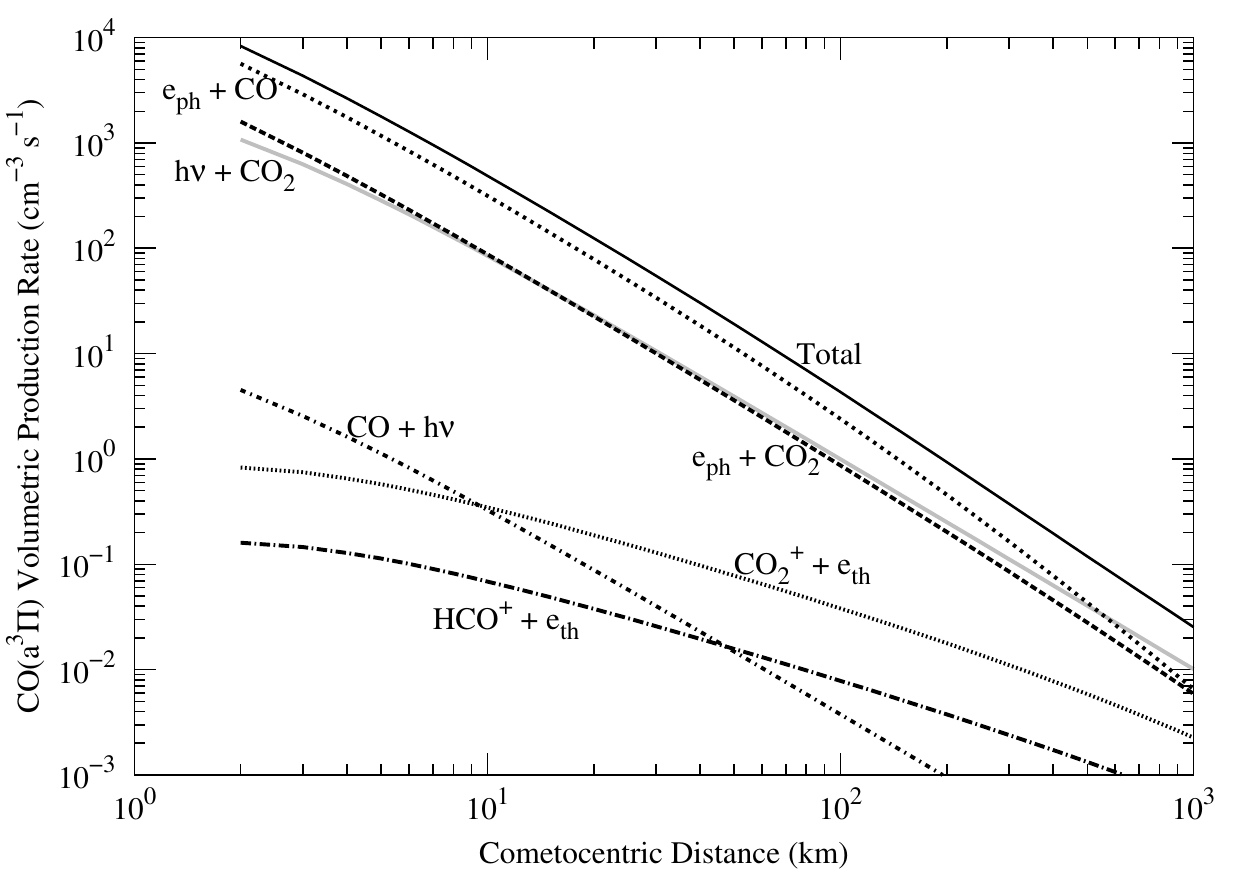}
   \caption{The calculated CO(\cam)  rate profiles  in
    comet \cg\ for an \wat\ out gassing  rate of  10$^{28}$
   s$^{-1}$ and for 5\% CO$_2$ and 5\% CO volumetric mixing ratios relative to water at 
   1.29 AU. 
   The photodissociation of \cod\ (gray curve) and suprathermal electron 
   impact on \cod\ (solid dashed curve)
   are producing CO(\cam) with nearly equal rates and both curves overlap.
    h$\nu$, e$_{th}$, and e$_{ph}$ stand for  photon, thermal electron,
     and suprathermal electron, respectively.
   \label{prd-cam}}
   \end{figure}

The modelled CO(\cam)  rate profiles  for different formation processes in the coma
of 67P are  shown in Figure~\ref{prd-cam}. The number of excited CO(\cam) molecules produced per unit volume per second is referred to as  volumetric production rate.  For
equal (5\% relative to water) CO$_2$ and CO volume mixing ratios relative to water 
 in the coma 
the major production sources of CO(\cam) is electron impact on CO. The electron
impact on \cod\ and photodissociation of \cod\ are next most dominant sources of 
CO(\cam).
The thermal electron recombination of HCO$^+$ and CO$_2^+$ ions and fluorescence
of CO are minor CO(\cam) production sources. Above 500~km the contributions from 
 photodissociative 
excitation of \cod\ and electron impact on CO and \cod\ are nearly equal. 
Since the lifetime of CO(\cam) is short \citep[$\sim$3~ms,][]{Gilijamse07}, most of the 
excited molecules decay to ground state by spontaneous emission.
Hence the radiative decay is 
the major loss source of this excited state.  Other loss processes, such as collisional 
quenching and ionization by photons and photoelectrons, are smaller compared to 
radiative decay by several orders of magnitude. The number of species de-excite to ground state
 per second
 by various loss mechanisms is referred to as 
loss rate.

\begin{figure}[h]
    \centering
   \includegraphics[scale=1]{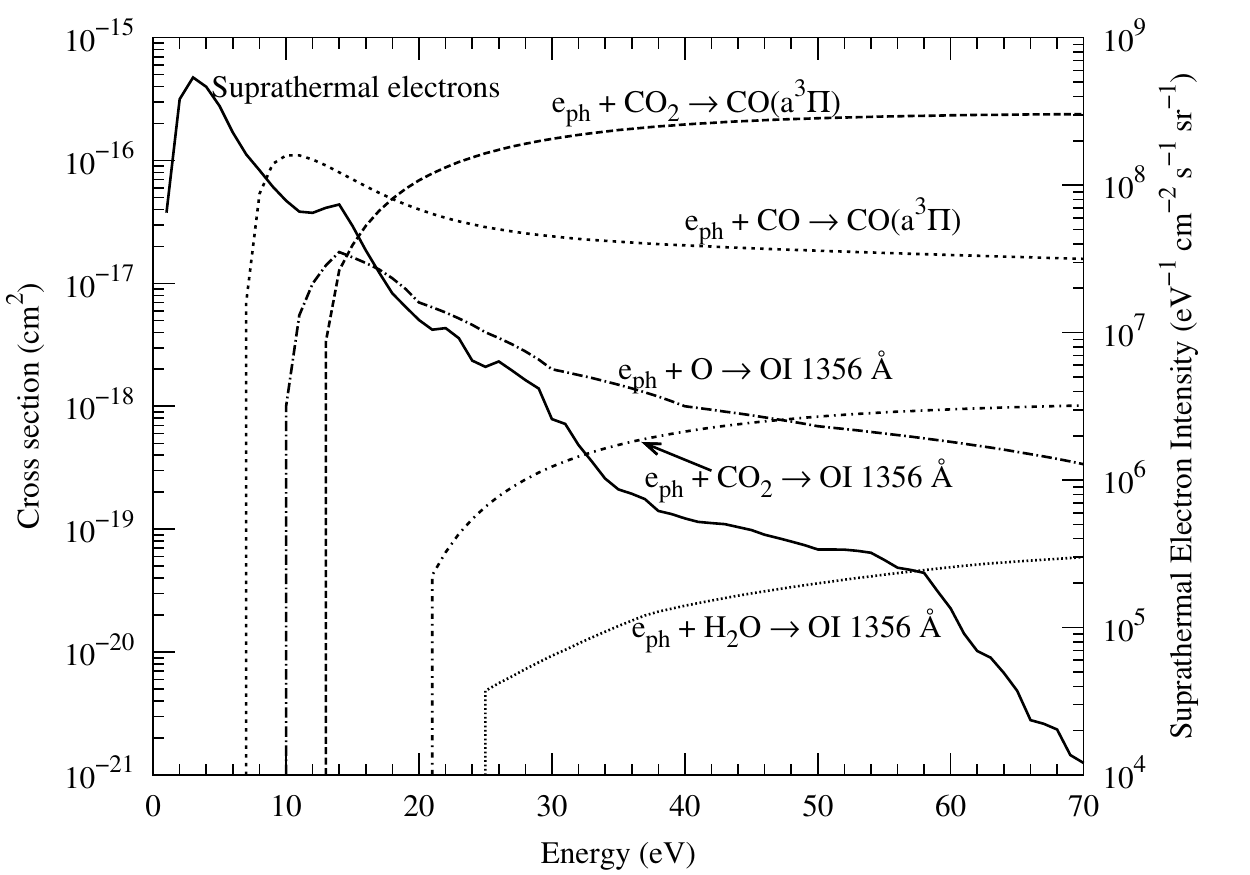}
    \caption{Cross sections for electron impact excitation of CO(\cam) from CO and \cod, and 
    for [OI] 1356 \AA\ from atomic oxygen, \cod, and \wat. 
    Calculated suprathermal electron intensity at cometocentric distance of 10~km is also shown 
    with magnitude on right side y-axis. The cross section for the formation of [OI] 1356~\AA\
    emission line from electron impact dissociation of H$_2$O is estimated by taking the
    ratio of cross sections for OI 1304~\AA\ to [OI] 1356~\AA\ at 100 eV from 
    \cite{Makarov04}. e$_{ph}$ stand for suprathermal electron.
    \label{csc-eflux}}
  \end{figure}

The cross section for electron impact excitation of various excited states and the 
calculated suprathermal electrons intensity in 67P coma at 10~km radial distance is presented in 
Figure~\ref{csc-eflux}. The suprathermal electrons  intensity in the energy range between 
10--15~eV 
mainly determines the excitation rate of CO(\cam) through electron impact on CO 
bearing species. 


  \subsection{Formation and destruction  of  atomic oxygen and atomic carbon}
  
  \subsubsection{Atomic oxygen in $^3$P, $^1$D,  $^1$S, and $^5$S states}

\begin{figure}[h]
    \centering
    \includegraphics[scale=1]{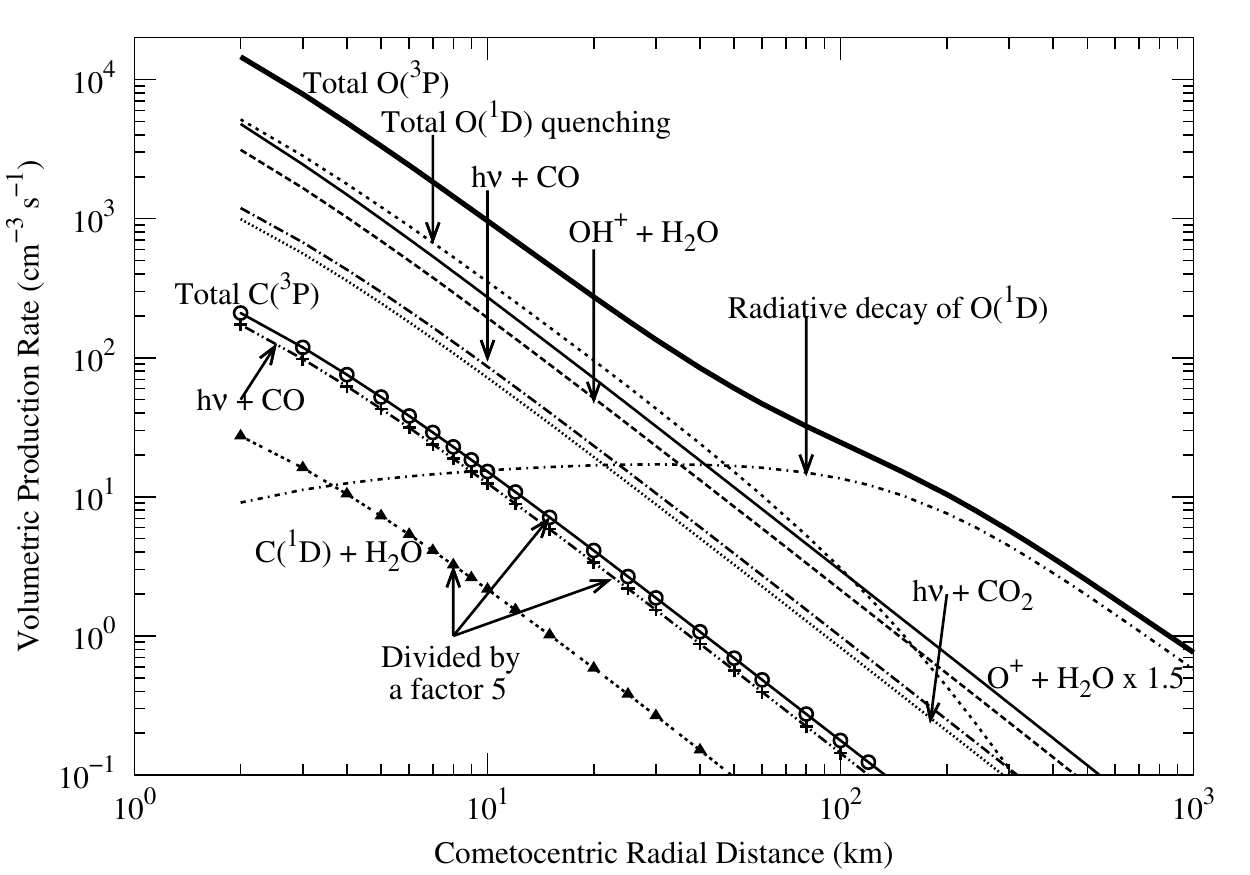}
    \caption{The calculated atomic oxygen and atomic carbon production rate profiles in 
      comet \cg\ with the water production rate of  10$^{28}$ 
     s$^{-1}$ for 5\% CO$_2$ and 5\% CO volume mixing ratios relative to
     water at 1.29 AU. The calculated atomic carbon production rate profiles, which are represented with symbols and curves, are divided by a 
     factor 5. The production rate profile of O($^3$P) through charge exchange between O$^+$ and \wat\ is multiplied by a factor 1.5. h$\nu$ stand for photon. \label{prod-ocg}}
    \end{figure}

 The modelled major production rate profiles of atomic oxygen and atomic carbon in 
  ground states are presented in Figure~\ref{prod-ocg}. Below 50~km radial distances, 
  various sources are contributing to the formation of \olg. 
  The production of  atomic oxygen  in ground state is mainly due to strong collisional 
  quenching of \old\ with water.  
  The next important sources of atomic oxygen are charge exchange of OH$^+$ and 
  O$^+$ ions with water.  Photodissociation of \cod\ and CO are 
  next important \olg\ production sources. Above 50~km  radiative decay of \old\ is the
  major source of atomic oxygen in ground state. 

\begin{figure}[h]
    \centering
    \includegraphics[scale=1]{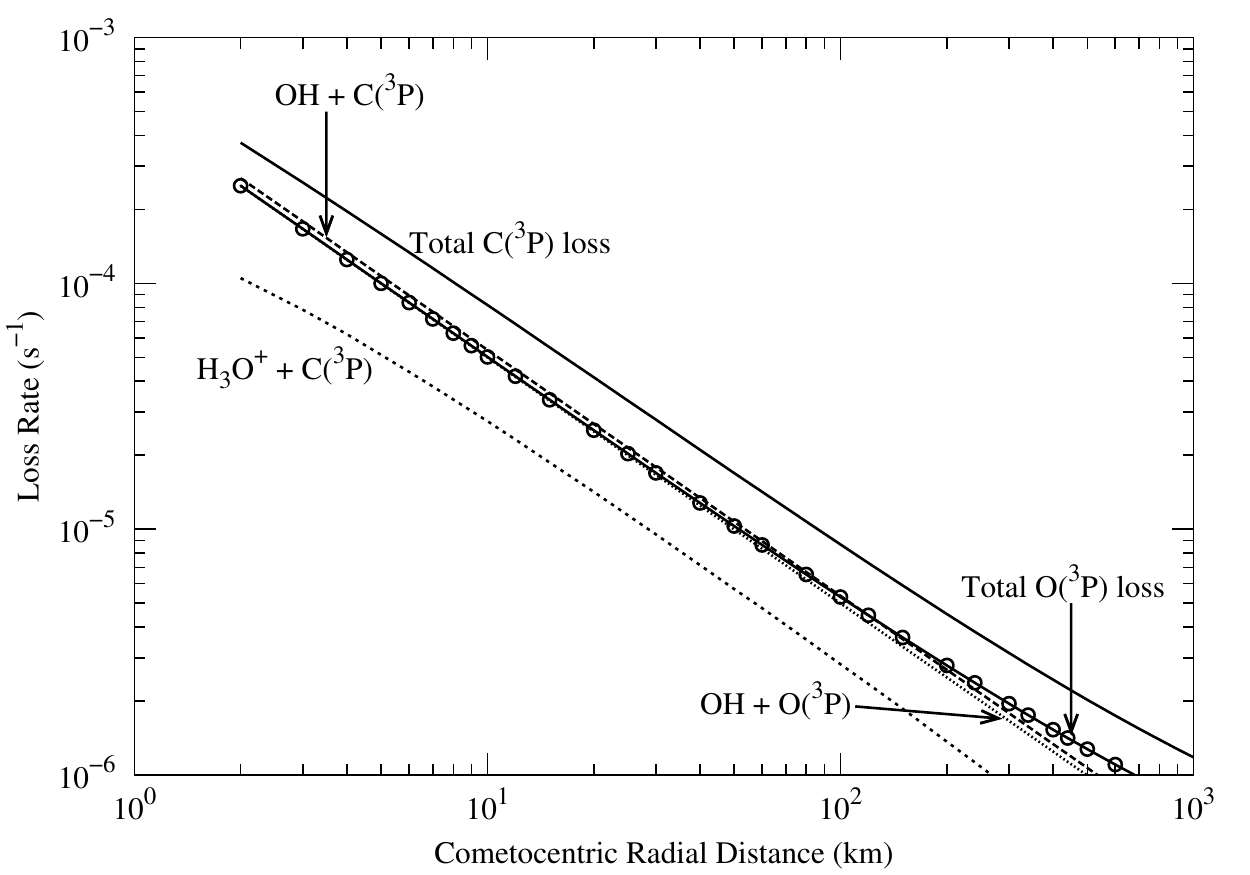}
    \caption{The calculated atomic oxygen and atomic carbon  loss  rate profiles in 
      comet \cg\ with the water production rate of  10$^{28}$ 
     s$^{-1}$ for 5\% CO$_2$ and 5\% CO volume mixing ratios relative to
     water at 1.29 AU.\label{los-ocg}}
    \end{figure}

 The calculated  major chemical loss rate profile  for \olg\  via different 
 destruction mechanisms are presented in Figure~\ref{los-ocg}. 
 The major loss process for the atomic oxygen is due to collisions with OH molecules 
 which yields atomic hydrogen and molecular oxygen. 
 The atomic oxygen can travel to
 large distances before getting lost in chemical reactions. Hence, we accounted transport loss
 by taking 1~km s$^{-1}$ as advection velocity.

\begin{figure}[h]
  \centering
  \includegraphics[scale=1]{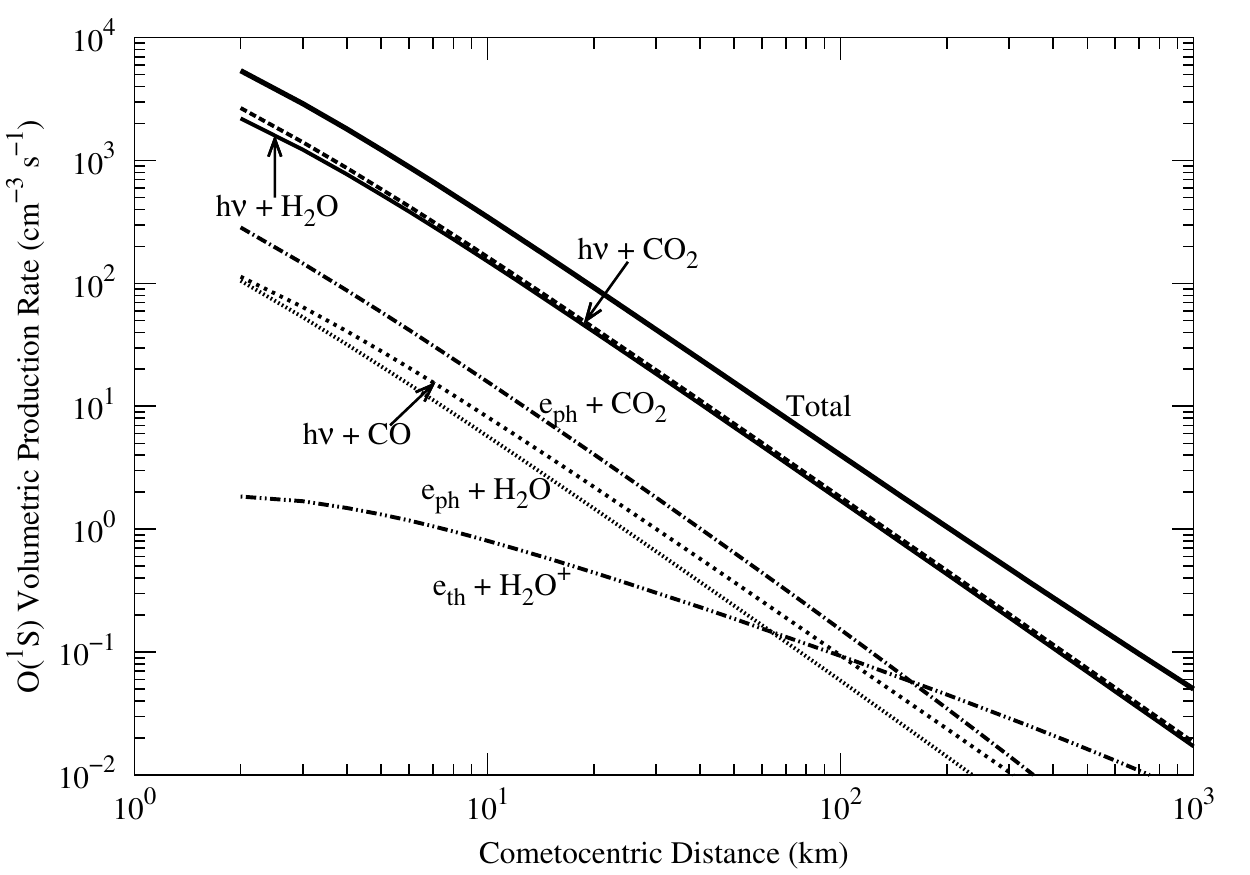}
  \caption{The calculated \ols\  rate profiles  in
  comet \cg\ for an \wat\ out gassing  rate of  10$^{28}$
  s$^{-1}$ and for 5\% CO$_2$ and 5\% CO volumetric mixing ratios relative to water at 
  1.29 AU. 
  h$\nu$, e$_{ph}$, and e$_{ph}$ stand for  photon, thermal electron,
  and suprathermal electron, respectively.
  \label{prd-o1s}}
  \end{figure}

  We have accounted for many \ols\ and \old\  formation and destruction processes in the
  coma as described in \cite{Bhardwaj12}. 
 The modelled major production rate profiles for O($^1$S)  in comet 67P  are shown in
 Figure~\ref{prd-o1s}. The photodissociation of \wat\ and \cod\ are equally important 
 sources in producing 
 \ols\ in the inner coma of comet 67P.  Below 100~km, suprathermal electron impact on \cod\ is 
 next important \ols\ source. 
  The photodissociation of CO and electron impact on \wat\
 and \cod\ are minor sources of \ols. Electron recombination of H$_2$O$^+$
 ion is a minor source of \ols\ in the inner coma 
  whereas its contribution is significant at large ($>$10$^3$~km) radial distances. 
 
\begin{figure}[h]
   \centering
   \includegraphics[scale=1]{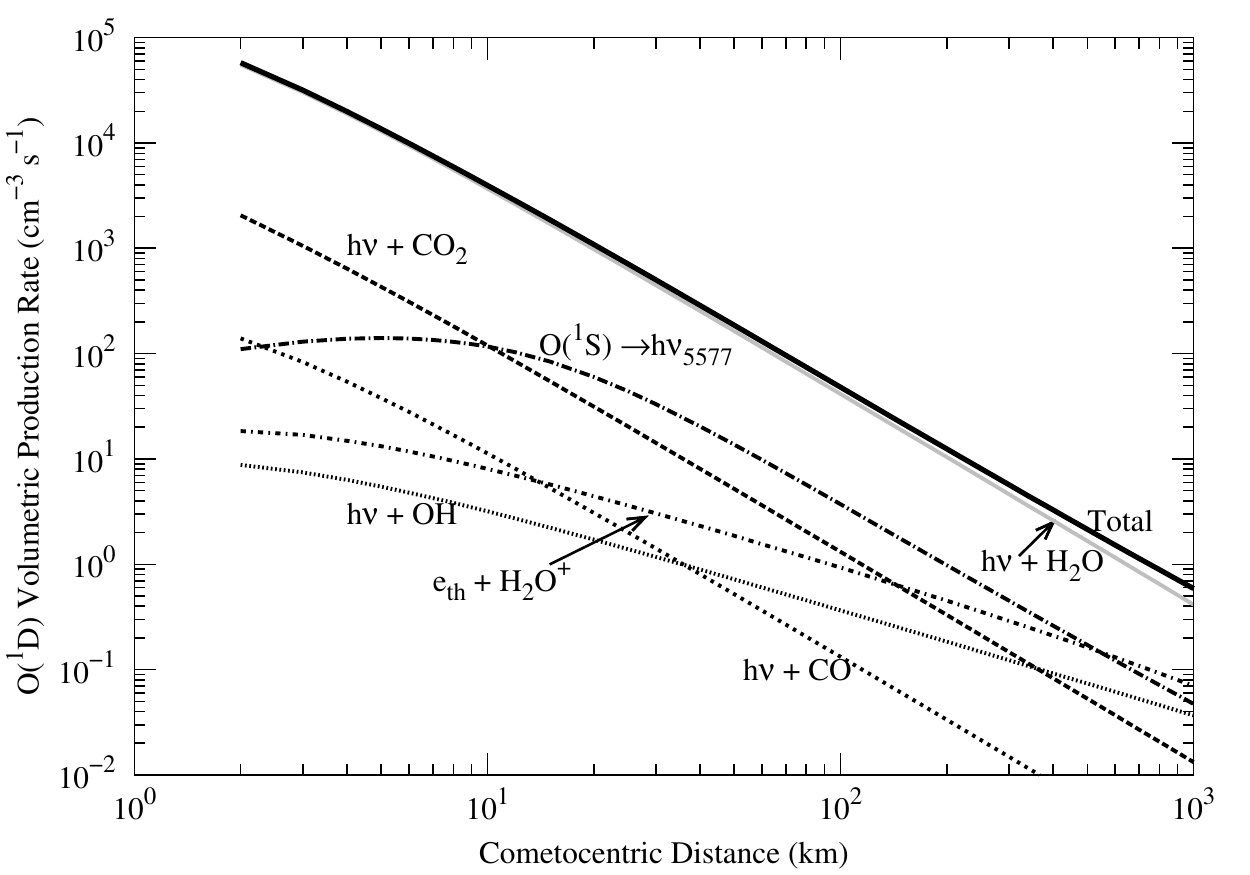}
   \caption{The calculated \old\  rate profiles in 
    comet \cg\ with the water production rate of  10$^{28}$ 
   s$^{-1}$ for 5\% CO$_2$ and 5\% CO volume mixing ratios relative to
   water at 1.29 AU. h$\nu$ and e$_{ph}$ stand for photon and 
   thermal electron, respectively.  \label{prd-o1d}}
   \end{figure}

 The calculated \old\ production rates profiles for different formation processes
 are presented in Figure~\ref{prd-o1d}. The photodissociation of \wat\ is a 
 dominant source of \old\ throughout the inner coma. Contribution from other \old\
 formation processes is minor  ($<$5\% to  the total). At large radial distances 
 ($>$~10$^3$~km) the
 contributions from dissociative recombination of H$_2$O$^+$, radiative decay of \ols\
  and photodissociation of 
 OH become significant in the formation of \old.

\begin{figure}[h]
  \centering
  \includegraphics[scale=1]{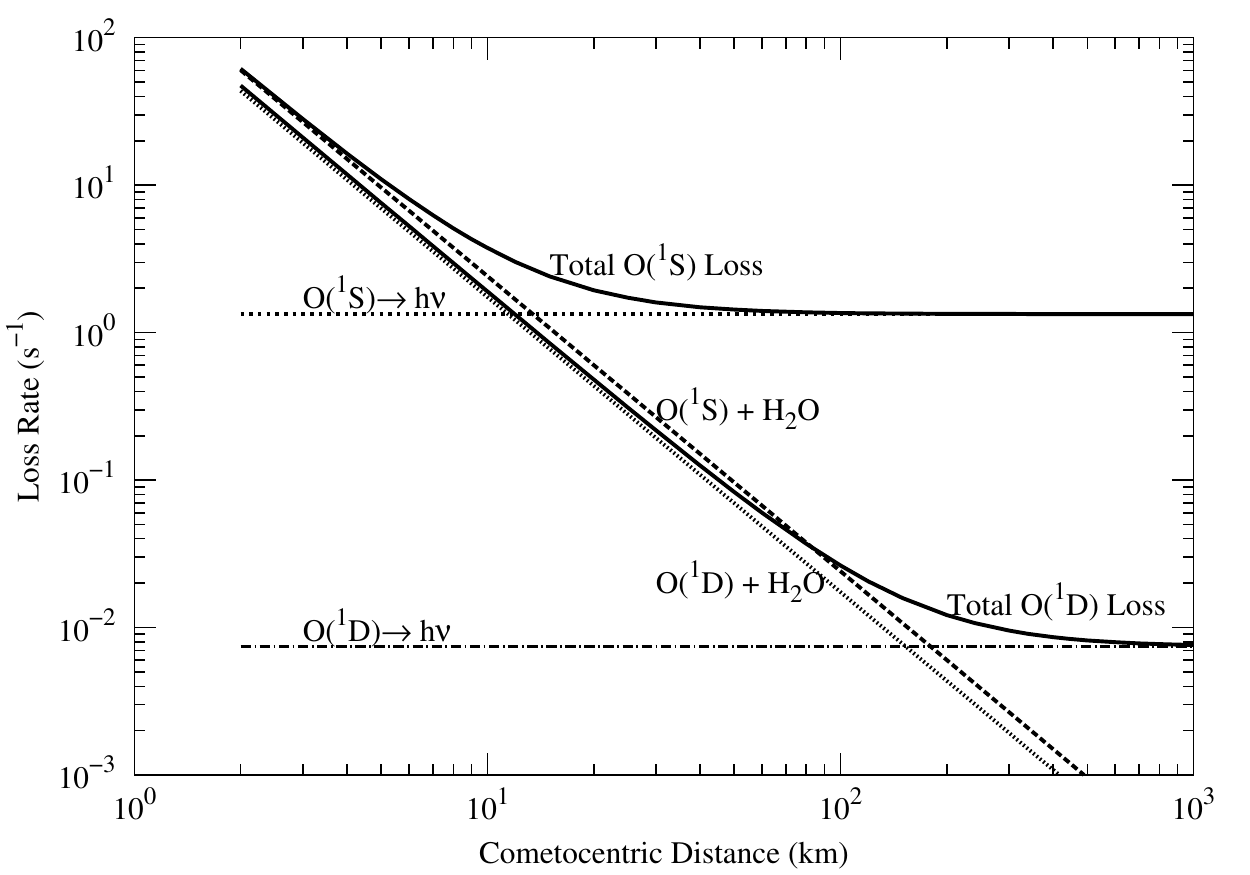}
  \caption{The calculated \ols\ and \old\ loss  rate profiles in 
    comet \cg\ with the water production rate of  10$^{28}$ 
   s$^{-1}$ for 5\% CO$_2$ and 5\% CO volume mixing ratios relative to
   water at 1.29 AU.  h$\nu$ stand for photon.\label{los-o1ds}}
  \end{figure}

 The modelled destruction rate profiles of \ols\ and \old\ are shown in
 Figure~\ref{los-o1ds}. The \ols\ and \old\ atoms are strongly quenched by \wat\
 up to radial distances of $\sim$10 and $\sim$200 km, respectively. Above these radial
 distances the radiative decay, which leads to forbidden visible emission lines, is the major
  loss process for the \ols\ and \old. Collisional quenching of \ols\ and \old\
  by other neutrals is  smaller compared to \wat\ quenching  by several orders of 
  magnitude, hence these processes are not shown in the figure.

\begin{figure}[h]
   \centering
   \includegraphics[scale=1]{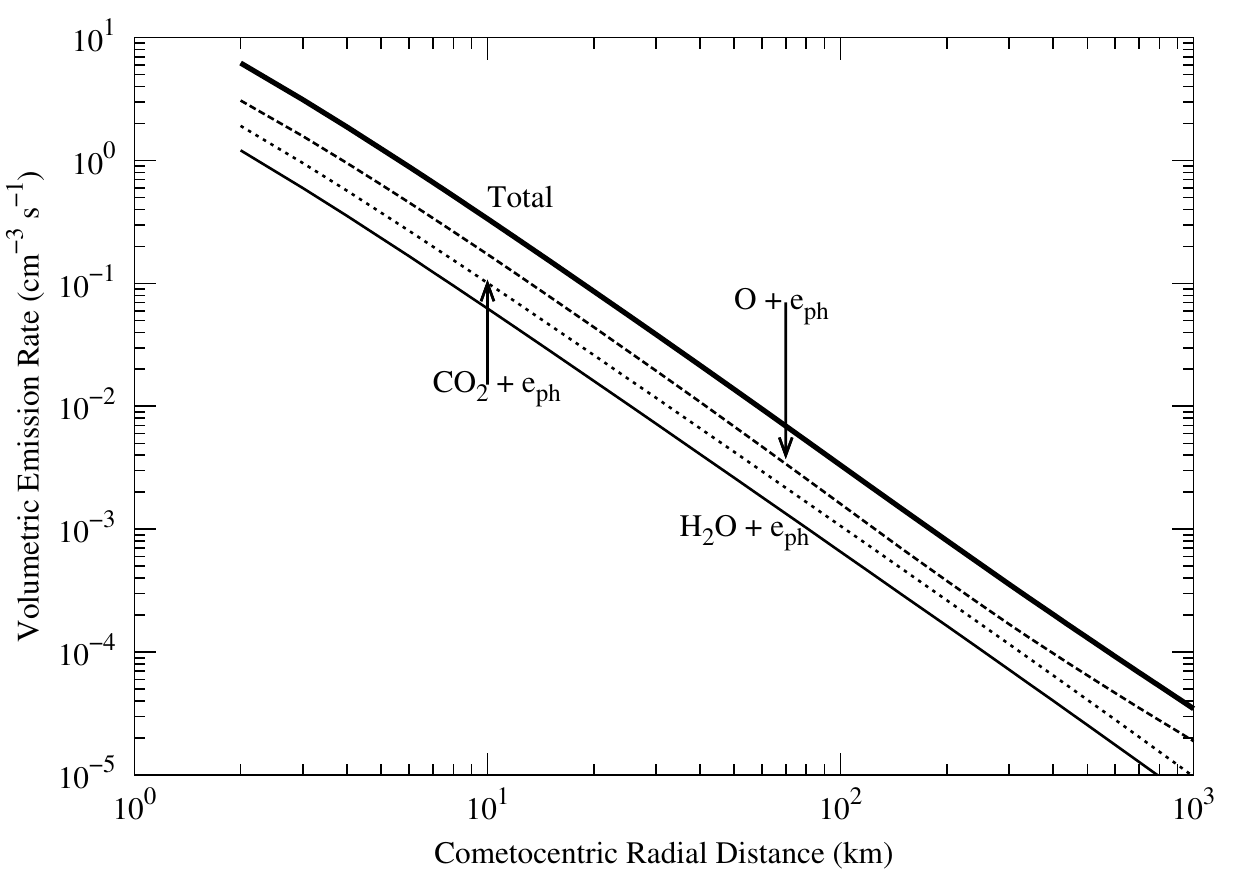}
   \caption{The calculated [OI] 1356~\AA\ emission rate profiles in 
     comet \cg\ with the water production rate of  10$^{28}$ 
    s$^{-1}$ for 5\% CO$_2$ and 5\% CO volume mixing ratios relative to
    water at 1.29 AU. e$_{ph}$ stand for suprathermal electron. \label{prod-oemis}}
   \end{figure}

 The production of atomic oxygen in $^5$S state yields [OI] 1356~\AA\ emission line
 via immediate decay to ground state \citep[lifetime is 185 $\mu$s,][]{Johnson72}.
 The calculated [OI] 1356~\AA\ emission rates are presented in Figure~\ref{prod-oemis}. 
 Electron impact on atomic oxygen is the major production source for [OI] 1356~\AA\ 
 emission followed by electron impact on \cod\ and H$_2$O.

\subsubsection{Atomic carbon in $^3$P and $^1$D states} 
 In case of atomic carbon  formation the photodissociation of CO is the major source of
\clg\ as shown in Figure~\ref{prod-ocg}. Collisional quenching of \cld\ is next important 
source of atomic carbon in ground 
state. All other production processes described in \cite{Bhardwaj96} contribute little 
($<$5\%) to the total. The loss of atomic
carbon is mainly  due to collisions with OH which yields atomic hydrogen and CO. 
The next main loss source is due to collisions with H$_3$O$^+$ which leads to the 
formation of 	HCO$^+$ and H$_2$. The model accounts for transport of atomic carbon in $^3$P and $^1$D sates  with an advection velocity of 1 km s$^{-1}$. Transport is the major loss 
processes for atomic oxygen and atomic carbon compared to the total loss due to 
chemical reactions.  
        
\begin{figure}[h]
       \centering
       \includegraphics[scale=1]{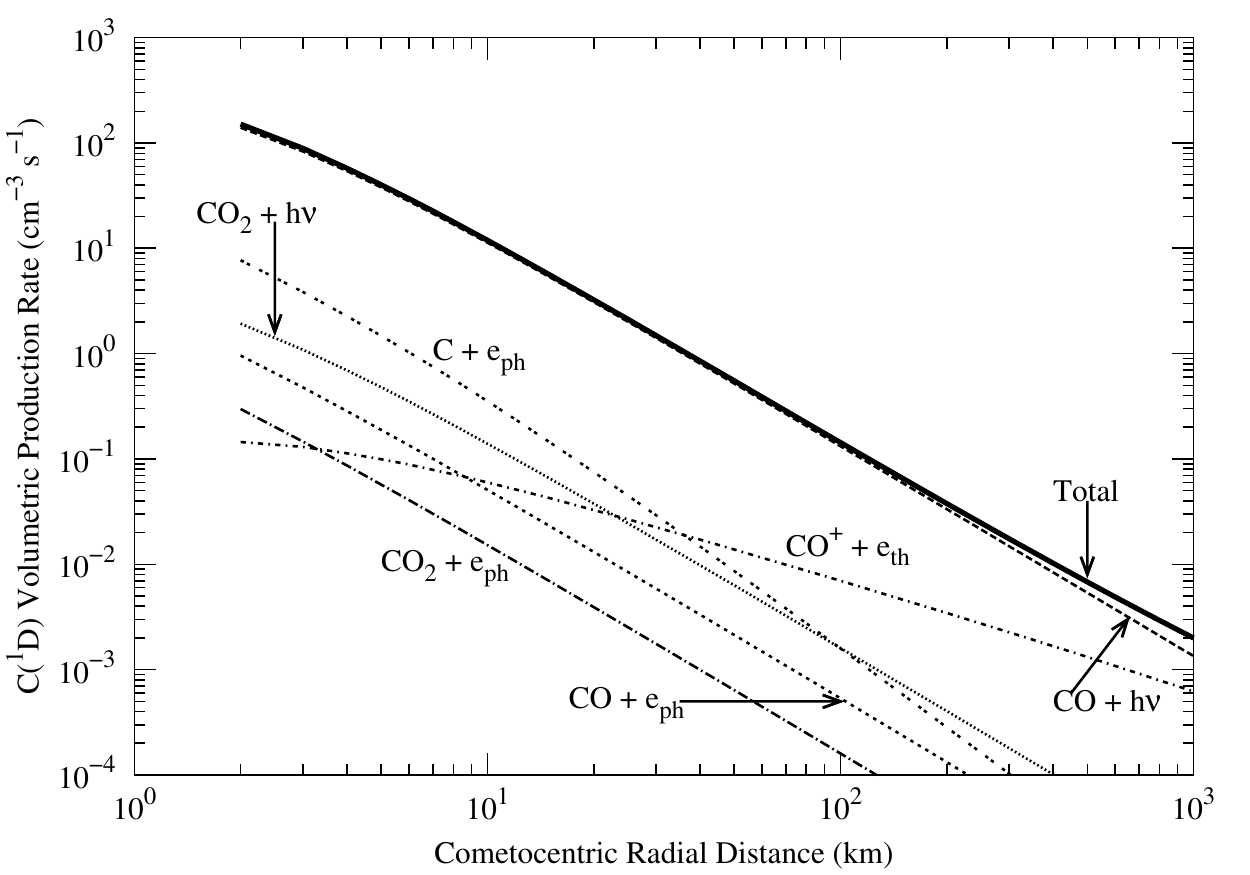}
       \caption{The calculated C($^1$D) production   rate profiles in 
         comet \cg\ with the water outgassing rate of  10$^{28}$ 
        s$^{-1}$ for 5\% CO$_2$ and 5\% CO volume mixing ratios relative to
        water at 1.29 AU. 
      h$\nu$, e$_{ph}$, and e$_{ph}$ stand for  photon, thermal electron,
       and suprathermal electron, respectively.
         \label{prod-c1d}}
       \end{figure}

\begin{figure}[h]
    \centering
    \includegraphics[scale=1]{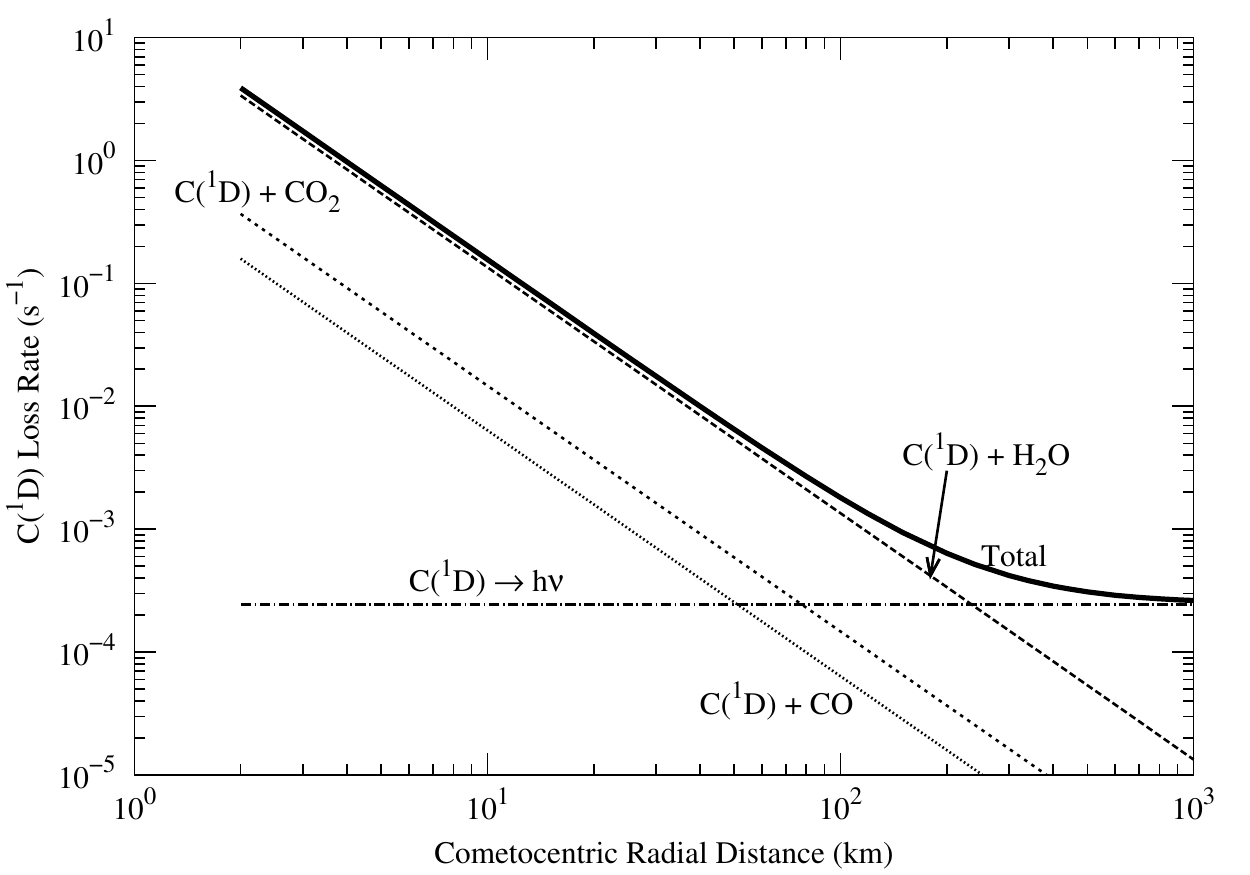}
    \caption{The calculated C($^1$D) loss  rate profiles in 
      comet \cg\ with the water production rate of  10$^{28}$ 
     s$^{-1}$ for 5\% CO$_2$ and 5\% CO volume mixing ratios relative to
     water at 1.29 AU. h$\nu$ stand for photon;\label{los-c1d}}
    \end{figure}

 The calculated formation rates of  metastable \cld\ atom via different production 
 processes are presented in Figure~\ref{prod-c1d}.  The major formation mechanism for
 \cld\ is photodissociation of CO. At large radial distance (10$^3$ km) dissociative 
 recombination of CO$^+$ ion is also an important source of \cld. Other formation 
 reactions are smaller compared to photodissociation of CO by more than an order of
 magnitude. The modelled  \cld\ loss rates  presented in Figure~\ref{los-c1d} show that 
 the collisional
 quenching with water is the dominant loss process up to 300 km radial distance, above 
 which radiative decay takes over. Collisional quenching of CO and \cod\ are relatively
 less significant \cld\ loss processes.  
  
\subsection{Calculation of emission intensities  along the projected distance}
\begin{figure}[h]
 \centering
 \includegraphics[scale=1]{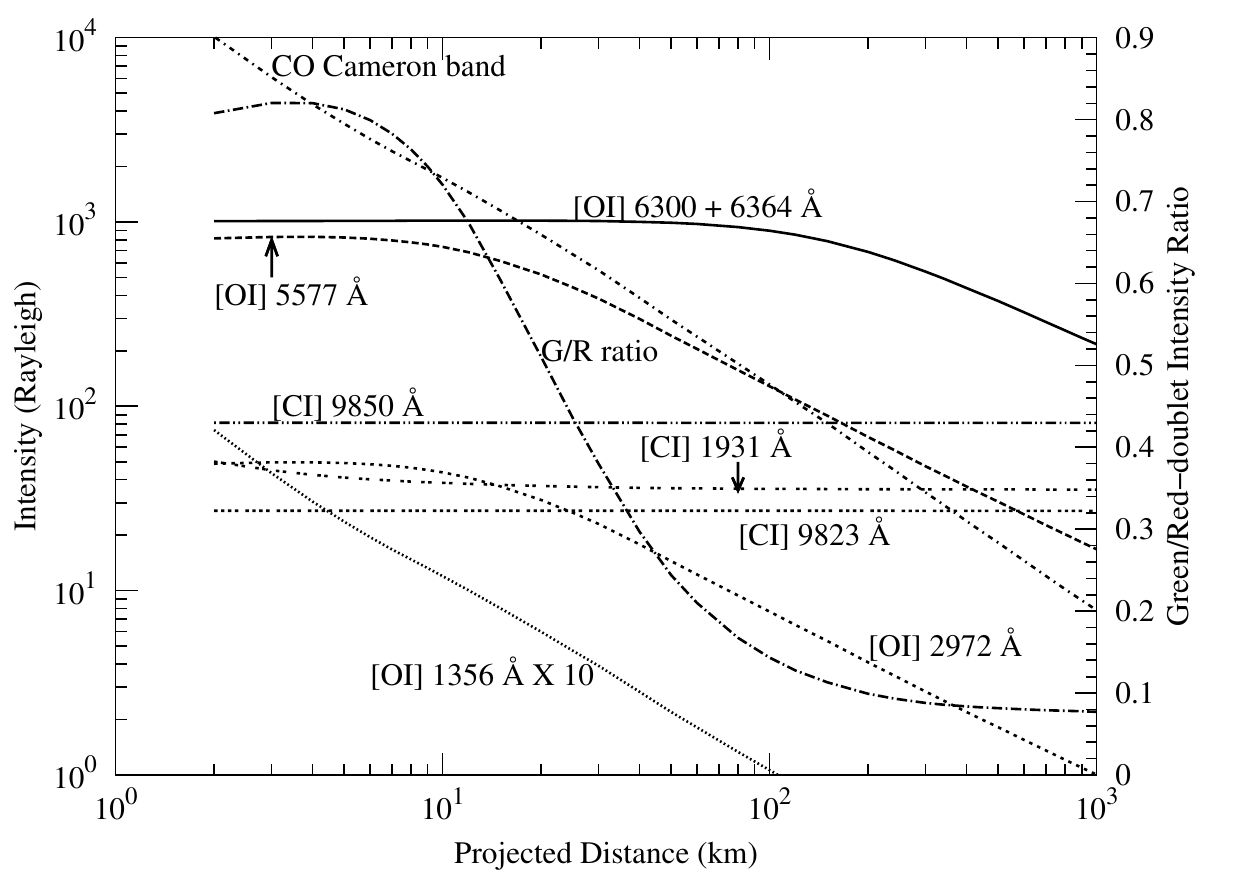}
 \caption{The calculated various emission 
 intensities for 5\% CO$_2$ and 5\% CO volume mixing ratios relative to water in comet \cg\
 with water production rate 1 $\times$ 10$^{28}$ s$^{-1}$ as a function of projected 
 distance at 1.29 AU.
  The calculated \grat\ values are shown on the right Y-axis. The [OI] 1356~\AA\ 
 emission line profile is multiplied by a factor 10. One Rayleigh = $\frac{10^6}{4\pi}$ 
 photons cm$^{-2}$ s$^{-1}$ sr$^{-1}$.
 \label{inten-emis}}
 \end{figure}

The radial emission rate profiles are integrated for each emission line along the line-of-sight perpendicular to the Sun-comet direction at different radial distances  to obtain  limb brightness profiles. 
The model calculated intensity profiles, as a function of projected distance, for CO
 Cameron band, atomic oxygen ( [OI] 6300+6364~\AA, 5577~\AA, 2972~\AA, and 1356~\AA) 
 and atomic carbon emissions ([CI] 1931, 9823, 9850~\AA) are shown in
Figure~\ref{inten-emis}. Among the calculated emission intensities CO Cameron band 
emission peaks close to the nucleus ($<$20 km). 
The calculated intensity profiles of 
5577~\AA\ and red-doublet (6300+6364~\AA)  are flat up to radial distances 20 
km and 200 km, 
respectively, due to strong collisional quenching of \ols\ and \old\  with \wat\ in 
the inner coma. The calculated \grat\ as a function of projected distance is also presented
in the same figure on the right Y-axis. Since lifetime of metastable \cld\ is large (4080 s),
the collisional quenching with water makes [CI] 1931, 9850 and 9823~\AA\ emission profiles  flat up to
1000 km.  The  [OI] 1356~\AA\ line is the weakest emission among the calculated  
emissions (presented in Figure~\ref{inten-emis}  after
multiplying a factor 10).

{Model calculated intensity profiles, 
when comet was at 3~AU heliocentric distance,  
are presented in Figure~\ref{inten-emis3au} as a function of projected distance.
 Very close to the nucleus surface the oxygen
 red-doublet, green line, CO Cameron, and [OI] 1356 \AA\ emissions are intense. 
 Since neutral gas production rate is low (5 $\times$ 
 10$^{25}$ s$^{-1}$) at 3~AU the calculated intensity profiles are decreased by 
two order of magnitude compared to those at perihelion. 
Due to high radiative lifetime ($\sim$110 s), the collisional quenching of \old\ is 
significant for radial distance
up to 20~km, which alters the \grat\ from 0.9 to 0.1. Inspite of having high mixing
ratio (15\%), the role of CO in determining the
oxygen visible emission intensities as well as in determining the \grat\ is insignificant. 
In the case of CO Cameron band,
most of emission intensity ($>$90\%) close to nucleus is mainly via electron
impact excitation of CO. For radial
distance higher than 50~km, the major ($\sim$50\%) source for CO(\cam)  is \cod\ via electron
impact and photodissociation excitation processes. The long radiative lifetime 
($\sim$4080~s) of 
\cld\ makes the atomic carbon emission intensity profile flat up to 1000~km.} 
 
\subsection{Effect of neutral composition on the calculation of emission intensities}

\subsubsection{Role of CO$_2$ and CO volume mixing ratios relative to water}
By varying the $\mu_w$(\cod) and $\mu_w$(CO) the 
contribution of different processes producing CO(\cam) is calculated at three different
projected distances. The calculations are presented in Table~\ref{tab:per-cam}
by varying $\mu_w$(\cod) and $\mu_w$(CO)  from zero to two, and then to five percent. In 
all these cases the contribution from
photodissociation of \cod\ and electron impact on \cod\ processes are nearly equal.
Keeping 1\% $\mu_w$(\cod) in the coma  we varied $\mu_w$(CO) between 1 and
 5\%. In this case
the contribution of photodissociation of \cod\ and electron impact on \cod\
 in the inner coma is $<$15\%,
whereas electron impact on CO is about 65--90\%. The contribution from other
chemical reactions, such as electron recombination of HCO$^+$ and CO$_2^+$ ions
contribute less than 10\% to the formation of CO(\cam). 

When $\mu_w$(\cod) is increased to 2\%, the contribution from electron
impact on CO is reduced to 50--80\%. About 10--20\% of CO(\cam) is
produced from photodissociation of \cod. By increasing $\mu_w$(\cod) 
to 5\%  the contribution of \cod, from both photodissociation and electron 
impact reactions, in producing CO(\cam) increases to a total of 30 -- 60\%.
In the case of  equal (5\%)  $\mu_w$(CO) and
$\mu_w$(\cod), the contribution from CO  is around 65\% and the
rest comes from \cod-associated reactions in the inner coma. 
 
Modelling results in Table~\ref{tab:per-cam} show that in the absence of \cod\ the main 
source for  
production of CO(\cam) is electron
impact of CO and other processes have negligible contribution to the total.
When CO is absent in the coma  electron impact on
\cod\ and photodissociation of \cod\ processes are producing CO(\cam) with
nearly equal contributions. In this case at 1000~km projected distances, 
the contribution of the thermal recombination of HCO$^+$ and CO$_2^+$ producing 
CO(\cam) is 15\%.

The calculated percentage contribution of different processes in the formation of 
\ols\ and \old\ are presented in Table~\ref{tab:per-o1sd}. The main processes 
controlling the formation of \ols\ is photodissociation of \wat\ and \cod. The 
photodissociation of \wat\ is the dominant source of \old\ production  in the inner 
coma.  Beyond 1000~km
radial distances, the photodissociation of OH, electron recombination of \wati, 
 and radiative decay of \ols\ are also important \old\ sources. 
The calculations presented in Table~\ref{tab:per-o1sd} show that below 100~km radial 
distance the formation of \old\ is mainly (80--90\%) through photodissociation of 
\wat.
Above these distances this contribution changes to around 50\%, while the  rest  
is via photodissociation of OH and dissociative recombination of \wati\ and radiative
decay of \ols.

In the case of \ols\ production, both photodissociation of \cod\ and \wat\ are
important formation processes in the inner coma. 
It is found that the role of CO photodissociation
is very small ($<$5\%) in the \ols\ production. Calculations presented in
Table~\ref{tab:per-o1sd} show that for 1\% of $\mu_w$(\cod), below 100~km projected distance, the contributions 
in the  formation of \ols\ are
 65--75\% from photodissociation of \wat, 15\%  from
\cod\ photodissociation,  and 20\% from other reactions. At 1000~km projected
distance the photodissociation of \wat\ is contributing around 45\% and
$\sim$45\% contribution mainly from dissociative recombination of \wati. In this
case the calculated \grat\ varies between 0.05 and 0.4 for less than 1000~km
projected distance. When we increased  $\mu_w$(\cod) to 5\% the contribution from 
both photodissociation of \wat\ and \cod\ is similar (30 to 45\%) for the projected
distances less than 1000~km. In this case \grat\ is found to vary between 0.07
to 0.7.
 
In the absence of \cod, the photodissociation of water and dissociative
recombination of \wati\ mainly controls the formation of \ols\ in the cometary
coma. In this case by changing the CO alone between 1 and 5\%, it is found that the
change in the calculated \grat\ profile is insignificant.
Assuming the absence of CO in the coma,  the calculated 
contributions are not changed from the previous cases whereas 
the calculated \grat\ profile is increasing linearly by increasing $\mu_w$(\cod).

We have calculated the [OI] 1356 
\AA\ emission intensity by varying $\mu_w$(\cod) and $\mu_w$(CO) between 1\% and 
5\%. The calculations show that electron impact on atomic oxygen is important (50\%) 
excitation process for [OI] 1356~\AA\ emission (see~Figure~\ref{prod-oemis}). Electron impact 
on  \cod\ is next important emission source of [OI] 1356~\AA\ line (30\% to the total). 
Below 50~km,  the 
formation of atomic oxygen is through collisional quenching of \old\ with water (35\%),
charge exchange between O$^+$ and OH$^+$ with water (45\%). Above this radial  
distance  75\% atomic oxygen is produced due to radiative decay of \old. 
The role of \cod\ and CO in producing atomic oxygen is less than 5\%.
Hence, by increasing the $\mu_w$(\cod) in the coma it is found that 30\% of this emission 
line intensity is increased only below 50~km radial distance. The role of CO in producing this emission line
is insignificant.    

The $\mu_w$(CO) can significantly influence the [CI] 1931~\AA\ emission 
intensity compared to that of \cod. By increasing the $\mu_w$(CO) from 1 to 5\% it is 
found that the intensity of this emission line also increases. 
The [CI] 1931~\AA\  is mainly through photodissociation of CO (75\%)  and \cod\ 
(20\%). The role of resonant scattering of \cld\ is significant (50\%) for  radial 
distances larger than 500~km. Similar effects also have been observed on 9850 and 9823 
\AA\ emission lines.

\subsubsection{Role of H$_2$O gas production rate}
The maximum gas production in this comet at perihelion for high activity case is about 1 
$\times$ 10$^{28}$ s$^{-1}$.  We have also done calculations for this comet 
by considering low activity case with water production rate of
 5 $\times$ 10$^{27}$ s$^{-1}$ and keeping the $\mu_w$(\cod) and $\mu_w$(CO) 
 equal 
 (5\%). By decreasing the gas production rate by a factor of 2
 it is found that the calculated emission intensities  are decreased by 30\%.  Similarly
 the collisional quenching radius for \ols, \old, and \cld\ also decreased by 30\% and is 
 moved towards the nucleus.  

\section{Discussion}
\label{discussion}
\subsection{Spectroscopic observations at comet 67P}
ALICE ultraviolet spectrometer
on-board Rosetta mission is designed to observe many emission
lines from 67P in the wavelength region 750 -- 2050~\AA\ \citep{Stern07,Feldman11}. 
This range overlaps with  part of  CO Cameron bands covering 1800--2600~\AA,
 [OI] 1356~\AA, and [CI] 1931~\AA\ emission lines \citep{Stern07,Feldman04}. 
By making limb scan observations  ALICE 
spectrograph can be used to derive  the spatial distribution of CO and/or \cod\ around 
67P nucleus. {ALICE can observe the shortward part of CO Cameron band emission in its longward 
limit where unfortunately the sensitivity is small \citep{Stern07}. 
Similarly [CI] 1931 \AA\ emission line also falls into the longer end of the ALICE spectral range.
It may however possible to detect it because of strong resonant fluorescence efficiency of C($^1$D) atom.
 The Rosetta onboard Optical, Spectroscopic, and Infrared Remote Imaging System (OSIRIS) is
 a scientific camera system with 12 discrete filters which is designed to observe 67P cometary coma
 over the wavelength rage 250--1000 nm \citep{Keller07}. OSIRIS can also map the release of certain 
 daughter species such as OH and [OI] based on observed emission intensities at 3090 \AA\ and
 at 6300 \AA, respectively.}

  Outcome of this study can be 
compared with on-board ROSINA mass spectrometers in-situ measurements of 
the neutral composition in the coma \citep{Balsiger07,Hassig15}. 
Space-based observations from Earth, such as from the Hubble Space Telescope (HST), can also
observe these ultraviolet emissions during Rosetta mission observation period. 
Several ground-based observatories have been observing comet 67P 
(\url{http://www.rosetta-campaign.net/planned-observations}) in visible and infrared 
regions to study the 
spatial distribution of various species. 
In this context the present
modelling work can provide better understanding of different processes governing
CO Cameron band, atomic oxygen and atomic carbon emission lines 
in comet 67P to derive parent neutral composition in the coma.

\subsection{Derivation of CO and \cod\ volume mixing ratios relative to \wat\ close to the nucleus}
By making several observations on different comets \cite{Tozzi98} demonstrated that  
there is a strong correlation  between  1931~\AA\ emission line intensity and CO 
column density.  The radiative 
decay of \cld\ to ground state yields 9823 and 
9850~\AA\ emission lines. By observing these
 emission lines  on Hale-Bopp, \cite{Oliversen02}  concluded that they can 
 be used as 
direct tracers of CO photodissociation in the cometary coma. The model calculations 
also
show that the major production source of \cld\ in the inner coma is mainly due to 
photodissociation of CO and the contribution from other production processes is smaller 
by an order of magnitude compared to the former (see Fig~\ref{prod-c1d}). 
The model calculated 1931, 9850, and 9823~\AA\  emission line profiles are flat up to 
1000~km projected distances due to strong collisional quenching of \cld\ with \wat. 
Since these atomic carbon  emission lines are mainly controlled by photodissociation of 
CO,  observed intensity profiles can be used to derive the CO gas production rate in 
the coma.  The calculated  emission intensity profile can be useful as a baseline 
prediction to constrain the CO mixing ratios in the coma  for the Alice observation of 
carbon emission lines which can be then compared with the ROSINA observations for
the same regions under similar solar illumination.

Measuring atomic oxygen visible emission line intensities is an important 
diagnostic tool in estimating the water production rate as well as to understand the spatial
distribution of \wat\ in the cometary coma \citep{Delsemme76, Delsemme79, Fink84, 
Schultz92, Morgenthaler01, Furusho06}. 
\cite{Decock15} analysed several ESO's VLT observed high resolution green and
 red-doublet emission line spectra  on 
various comets. 
\cod\ mixing ratios  are derived  in these comets by comparing the ESO VLT 
observations with modelled \grat\ profiles \citep{Decock15}. The model calculated \grat\ 
profile is 
presented on 67P in Figure~\ref{inten-emis}. By modelling green and red-double 
emission intensity  profiles on various comets at different heliocentric distances, 
\cite{Raghuram14} 
have shown that \grat\ 
value  increases linearly by increasing $\mu_w$(\cod) in the coma whereas the 
affect of  CO is minor in determining  either green or red-doublet emission intensities. 
Hence, the observed \grat\ profile on 67P can be used to constrain \cod\ mixing ratio 
in the coma. 

\begin{figure}[h]
 \centering
 \includegraphics[scale=1]{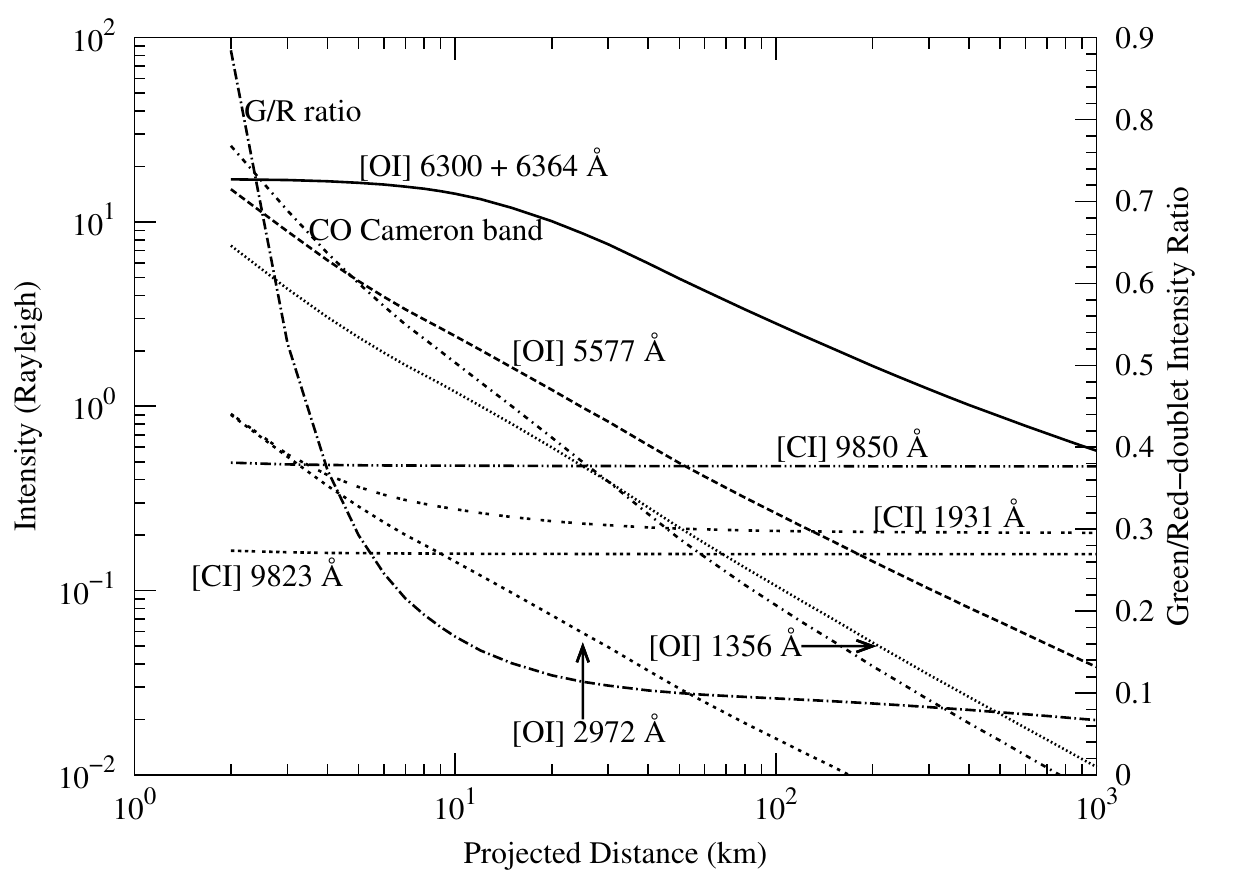}
 \caption{The calculated various emission 
 intensities for 5\% CO$_2$ and 15\% CO volume mixing ratios relative to water in comet \cg\
 with water production rate 5 $\times$ 10$^{25}$ s$^{-1}$ as a function of projected 
 distance at 3 AU. The calculated \grat\ values are shown on the right Y-axis. One Rayleigh = $\frac{10^6}{4\pi}$ 
 photons cm$^{-2}$ s$^{-1}$ sr$^{-1}$.
 \label{inten-emis3au}}
 \end{figure}

{When a comet is far away from the Sun (3 AU), 
it is expected to have higher CO and \cod\ volume mixing ratios which are species associated
with low sublimation temperatures \citep[e.g.,][]{Mumma11}. 
The calculations made at 3~AU heliocentric distance (see Fig.~\ref{inten-emis3au}), with 
mixing ratios 5\% \cod\ and 15\% CO, 
show that atomic oxygen red-doublet emission is the most intense
emission in the inner coma. This emission can be observed by Rosetta onboard 
OSIRIS instrument which can be subsequently used to derive water production rate. 
Unfortunately there are 
no filters on OSIRIS to measure [OI] 5577 \AA, [CI] 9823 \AA, and 
9850 \AA\ emission lines \citep{Keller07}.  The predicted oxygen red-doublet intensity along the
projected distance (Figures~\ref{inten-emis} and \ref{inten-emis3au}) could be 
useful in analysing OSIRIS visible spectra of 67P and subsequently deriving
H$_2$O distribution around the nucleus.}

\subsection{Constraining the  \ols\  yield for \wat\ at solar Ly-$\alpha$ wavelength}
The photon cross section for the formation of \ols\ from \wat\ has never been reported in the 
literature \citep{Huestis06}. In this model the formation of \ols\ from photodissociation of
\wat\  has been accounted by assuming 0.5\% yield for \wat\ at Ly-$\alpha$ wavelength
\citep{Bhardwaj12}.
The onboard ROSINA spectrometers can measure the \cod\ number densities  during this 
mission 
period at different radial distances in the coma.  
 By combining the observed \grat\ profile with onboard ROSINA CO$_2$
 measurements
it would be  possible to constraint the \ols\ average yield value at solar 
Lyman-$\alpha$.   
  The high-resolution spectroscopic observations, such as analysis of \cite{Decock15}, 
can provide 
information about collisional quenching of \ols\ and \old\ 
metastable states in the coma of 67P.  The observation of both green and red-doublet
 emission line widths and 
\grat\ profiles along with ROSINA measurements can solve the puzzle that green line is 
wider than either of red-doublet emission lines in comets.

\subsection{Derivation of suprathermal electron intensity close to the nucleus}

The modelling of  production rates of CO(\cam) has shown that suprathermal electron 
impact reactions mainly govern the CO Cameron band emission with contribution of 
around 75\% whereas \cod\ photodissociation contributes  about 25\% (see 
section~\ref{results_cam}). 
In the absence of CO, the
 electron impact on \cod\ is an equally important production source of CO(\cam) as 
 photodissociation of \cod\ (see Table~\ref{tab:per-cam}). This suggest that 
electron impact excitation mechanism
should be considered for the estimation of parent species production rates in the coma. 
With sufficient CO  ($\ge$3\%)  in the coma, the contribution from
electron impact reactions in producing this band emission close to the nucleus 
($<$100~km) is about 80\%. The excited state CO(\cam) is mainly populated in the coma by 
suprathermal electrons in the energy range between 10 and 15 eV (see Figure~\ref{csc-eflux}).
Since the major source for the production of CO(\cam)
is electron impact,  the observed CO Cameron band emission close to the nucleus would 
be suitable to track the suprathermal electron intensity \citep{Mcphate99} rather than \cod\ neutral 
density. 

The [OI] 1356~\AA\ emission line is an excellent tracer for electron impact processes in the 
coma. Modelling of electron impact excitation processes shows that this emission is mainly due 
to electron impact excitation of atomic oxygen followed by electron impact dissociative 
emission of \cod\ and \wat\ (see Figure~\ref*{prod-oemis}). However, the intensity of this emission 
 is weaker by three orders of magnitude compared to CO Cameron band emission.  
The electron impact excitation cross section for atomic oxygen producing 
 [OI] 1356~\AA\ emission line peaks at 15 eV whereas for \cod\ and \wat\ it is between
 30 and 60 eV.  The contribution from atomic oxygen is about 50\% and the rest is through
 \cod\ (30\%) and \wat\ (20\%). Hence, half of the observed  emission intensity profile
 is linked to the suprathermal electron intensity at 15 eV. 
 Recently ALICE observed several HI, OI, and CI emission near the cometary nucleus when comet was at 
 around 3 AU \citep{Feldman15}. The observation of OI 1356 \AA\ emission line intensity, which is varying between 
  $\sim$1.5 and $\sim$3 Rayleigh at 10 km projected distance, \citep{Feldman15}, is close to our predicted calculation ($\sim$1.5 Rayleigh, see Fig.~\ref{inten-emis3au}). Detailed analysis of this emission 
  line will be presented in the future work using the Rosetta measured neutral density 
  distribution around the nucleus.
 
The onboard Rosetta Plasma Consortium (RPC)/ Ion and Electron Sensor (IES) 
 is capable of measuring the electron 
energy spectra in the energy rage 1 eV/e to 22 keV/e  \citep{Burch07}. Since both CO 
Cameron band and [OI] 1356~\AA\ 
emission lines are governed mainly by electron impact excitation reactions, the observed 
emission intensities may be supportive for the IES  measured 
suprathermal electron intensity at around 15 eV. 

\subsection{Implication of molecular oxygen in determining the \grat}
Molecular oxygen is the major source for the production of \ols\ and \old\ in the terrestrial atmosphere.
 The recent discovery of O$_2$ in 67P's coma by ROSINA/DFMS \citep{Bieler15} demands the inclusion
  of O$_2$ in the model in order to calculate green and 
 red-doublet emission intensities.
 By including 4\% molecular oxygen with respect to water production rate,  and for the input conditions described in Section~\ref{model-inputs},
  the  \grat\ is found to be increase by around 20\% close to the nucleus ($<$20 km projected distance).  \cite{Bieler15} observed
 that the relative abundance of molecular oxygen ranges 1 to 10\% with respect to \wat\ production rate. Hence
 in order to determine CO$_2$ abundance based on the \grat\ the contribution from molecular oxygen should also
 be considered. In case of higher O$_2$ abundance in comets,
  the observed \grat\ can be significantly controlled by photodissociation of O$_2$ and may lead to underestimation of  CO$_2$  mixing ratio.
 
 \subsection{Parameters which can influence the predicted emission intensities}
 The estimated diamagnetic cavity on the sunlit side of this 
 comet at perihelion is around 30--40~km  \citep{Benna06,Hansen07,Koenders15}. 
 The extent of  diamagnetic 
 cavity depends on the gas production rate and solar wind conditions during comet 
 perihelion visit. Beyond this cavity, most of the ions are transported towards the tail side 
 due to solar wind interaction.
 The assessment of solar wind interaction on the emission intensities is beyond the scope 
 of this work but we would like to discuss the possible sources that can alter the emission 
 intensities. 
 Outside the diamagnetic cavity, the chemical lifetime of neutrals  can  be 
 significantly altered by charge exchange between solar wind ions and cometary species. 
 Hence, it is expected that the calculated intensities outside the diamagnetic cavity 
 can be changed based on the solar wind conditions during that time. 
 The electrons outside the diamagnetic cavity are primarily solar wind electrons or
 shocked solar wind electrons \citep{Reme91, Gringauz86,Gan90, Cravens91, Ip04}.
 The population of suprathermal electrons outside the diamagnetic cavity is a complex problem 
 due to admixture of solar wind electrons. However,  the radius of 
 collisional zone and diamagnetic cavity are subjected to the gas production rate and 
 solar wind conditions during the comet perihelion passage.

 For electron impact driven emissions, such as CO Cameron band and 
 [OI] 1356~\AA,   due to strong solar wind interaction 
 both neutral density and electron 
 population may change outside the diamagnetic cavity region, thus the observed emission intensities vary significantly.
 In this region the solar wind electrons may also contribute to the total emission 
 intensity \citep{Bhardwaj90,Bhardwaj96,Bhardwaj99a}.  However, the dissociative
 recombination CO-bearing ions to the total emission intensity  contribute little ($<$5\%), 
 whereas formation of atomic oxygen significantly (50\%), because of charge exchange 
 between O$^+$ and OH$^+$ with \wat.
 We do not expect the radiative decay and collisional quenching of \old\ can be altered 
 significantly due to solar wind interaction. 
 
  The evolution of cometary ionosphere around 67P nucleus has been monitored by 
 Rosetta Plasma Consortium and Rosetta Orbiter Spectrometer for Ion and Neutral Analysis (ROSINA) instruments. The recent observations of ROSINA/Double Focussing Mass Spectrometer (DFMS), RPC/Ion and Electron Sensor (IES), and
 RPC/Ion Composition Analyzer (ICA),
  when the comet was beyond 2 AU, have shown that due to low outgassing rate no contact surface is formed and 
 most of the solar wind has directly accessed the 67P's nucleus, though the plasma close to the comet is dominated
 by cometary water ions \citep{Fuselier15,Broiles15,Nilson15}.
 \cite{Clark15} found that the suprathermal electrons
  are accelerated to 
 several hundreds of eV. The high energetic solar wind charged particles may also be involved in producing 
 the excited atomic and molecular states discussed here.

The formation \ols, \old, and \cld\ are mainly due to photochemical reactions.  The 
contribution from  ions and thermal electron recombination reactions for the  
inner coma is very small ($<$5\%). Hence, we do not expect the predicted oxygen visible 
and [CI] 1931~\AA\ line emission intensities to change due to solar wind interaction for 
the inner coma unless the radial distribution of \wat\ is changed. 

In this model we have accounted for  main parent oxygen- and carbon-bearing species
to compute the emission intensities. However, the contribution from photodissociation 
and electron impact of other minor species is also possible. In case of atomic oxygen 
visible emissions, the dissociation 
of other oxygen-bearing  species, such as HCOOH or H$_2$CO, are unlikely to be the 
parent because they cannot decay fast enough to produce \old\ and \ols\ 
\citep{Festou81}. However,  in the case of carbon emissions there could be  an 
involvement from other carbon-bearing species, such as hydrocarbons. Since the major 
processes governing these emissions are via photochemical reactions, the role of electron
temperature for the inner coma is not significant ($<$5\%). The model calculations are done
for gas phase so scattering of solar photons by dust grains could be a significant factor in 
governing these emission intensities. 

The recent ROSINA/DFMS  observations on 67P show that the cometary
coma contains a variety of species with
heterogeneous distribution which varies with time and latitude \citep[e.g.,][]{Hassig15,Leroy15}.
 The Microwave Instrument on the Rosetta Orbiter (MIRO) mapped around 67P's nucleus when 
 it was at 3.4 AU \citep{Biver15,Lee15}. The water column density in the 
inner coma (within 3 km from the nucleus) is found to vary even by two orders of magnitude. 
\cite{Kuti15} further investigated the heterogeneity of 
67P's coma by measuring various major (\wat, CO$_2$, and  CO) and minor
(HCN, CH$_3$OH, CH$_4$ , and C$_2$H$_6$) volatile species using ROSINA/DFMS.
 Our calculated emission intensities  
 may change significantly  due to variable neutral densities around comet 67P. 
Future work will include the use of in-situ measured neutral densities from ROSINA sensors
to drive model calculated emission intensities. Results could then be compared 
 with ALICE and OSIRIS observations as well as ground-based 
 observations.

	 \section{Conclusions}
 \label{conclusions}
Rosetta-remote and Earth-based spectroscopic observations, combined with
 modelling of comet 67P, offer a unique opportunity to assess the main production and 
 destruction  processes governing various  forbidden visible and ultraviolet emissions
  as the comet gets closer to the Sun. 
 The combined analysis applied to Rosetta remote and in-situ observations could be used as a 
 ground truth for the interpretation of Earth-based observations on 67P cometary coma.
 The model calculations suggest that 	the electron 
impact reactions are the dominant  sources in producing CO Cameron band 
emission. Hence, the observed 
 CO Cameron emission intensity close to the cometary nucleus can be used to track the 
 suprathermal electron intensity  in the 
 energy range 10 to 15 eV close to the nucleus. 
The observed \grat\ away from the collisional zone can be used to confirm the parent 
oxygen species producing these emissions. Measurement of the \grat\ close to the comet
 as a function
of projected distance can be used to constrain the $\mu_w$(\cod). Presence of high mixing ratio
of molecular oxygen can affect the \grat\ significantly which may lead to underestimation of $\mu_w$(\cod).
 The observation of 
{[OI] 1356 \AA\ can give a clear indication of the role of electron 
impact processes in the coma, while}  
[CI] 1931 \AA\ emission is a good tracer to probe CO distribution near the nucleus. Both 
Cameron band and atomic oxygen emission observations are useful to assess  \wat, \cod, 
and CO volume mixing ratios in the coma and to understand the spatial distribution and 
their time evolution in comet 67P. {The quantitative assessment
of different excitation processes is essential to study  the evolution of the
chemistry in the inner cometary coma
with the increasing neutral gas production rate.} 

\begin{acknowledgements}
S.R.  and M.G. acknowledge the support of the Science and Technology Facilities 
Council (STFC) of UK through the Consolidated Grant ST/K001051/1 to Imperial College London. The work of A.B. is supported by Indian Space Research Organization.
\end{acknowledgements}

\newcommand{\noopsort}[1]{}

\clearpage

\begin{sidewaystable}[h]
\caption{The calculated contribution of different production processes 
producing CO(\cam) in comet 67P for different \cod\ and CO volume mixing ratios relative to water. 
\label{tab:per-cam}}
\begin{tabular}{cclllllllllllll}
\toprule
\multicolumn{2}{p{4cm}}{\centering Volume mixing ratios relative to water (\%)} & 
 \multicolumn{12}{c}{\centering Percentage contribution
at different projected distances (km)} \\[-22pt]
 & & \multicolumn{3}{c}{h$\nu$ +\cod} & \multicolumn{3}{c}{e$_{ph}$ + \cod} &  
 \multicolumn{3}{c}{e$_{ph}$ + CO} 
&  \multicolumn{3}{c}{Others\footnotemark[2]}  \\
\midrule
\cod & CO  & 10 & 10$^2$ & 10$^3$ & 10 & 10$^2$ & 10$^3$& 10 & 10$^2$ & 
10$^3$ & 10 & 10$^2$ & 10$^3$ \\ 
\cmidrule(lr){3-5} \cmidrule(lr){6-8} \cmidrule(lr){9-11} \cmidrule(lr){12-14}
1 &1  & 14.2 & 14.5 & 13.9 & 15.8 & 15.4 & 14.7 & 69.7 & 68.5 & 65.4 &  0.2 &  1.6 & 
6.0 \\
1 &2  &  9.1 &  9.3 &  9.1 &  9.6 &  9.5 &  9.1 & 81.1 & 80.1 & 77.5 &  0.2 &  1.1 &  4.3 \\
1 &5  &  5.0 &  5.2 &  5.1 &  4.8 &  4.7 &  4.6 & 90.1 & 89.3 & 87.2 &  0.1 &  0.8 &  
3.1\\[5pt]
2 &1  & 21.6 & 21.9 & 20.6 & 24.4 & 23.6 & 22.1 & 53.6 & 52.2 & 48.8 &  0.4 &  2.3 &  
8.5\\ 
2 &2  & 15.2 & 15.5 & 14.8 & 16.3 & 15.9 & 15.1 & 68.3 & 66.9 & 63.7 &  0.3 &  1.7 &  
6.4\\
2 &5  &  9.0 &  9.3 &  9.1 &  8.7 &  8.6 &  8.3 & 82.1 & 81.0 & 78.2 &  0.2 &  1.1 &  4.5\\ 
[5pt]
5 &1 & 31.4 & 31.5 & 28.9 & 36.5 & 34.8 & 31.7 & 31.6 & 30.4 & 27.7 &  0.5 &  3.3 & 
11.7\\
5 &2  & 25.2 & 25.5 & 23.8 & 28.0 & 26.9 & 24.9 & 46.4 & 44.9 & 41.6 &  0.4 &  2.7 &  
9.8\\
5 &5  & 17.3 & 17.7 & 16.9 & 17.4 & 16.9 & 15.9 & 65.0 & 63.4 & 59.9 &  0.3 &  2.0 &  
7.3 &\\[5pt]
0 &1  &  0.0 &  0.0 &  0.0 &  0.0 &  0.0 &  0.0 &100.0 & 99.8 & 99.0 &  0.0 &  0.2 &  1.0\\
0 &2  &  0.0 &  0.0 &  0.0 &  0.0 &  0.0 &  0.0 &100.0 & 99.8 & 98.9 &  0.0 &  0.2 &  1.0\\
0 &5  &  0.0 &  0.0 &  0.0 &  0.0 &  0.0 &  0.0 &100.0 & 99.7 & 98.7 &  0.0 &  0.3 &  
1.3\\[5pt]
1 & 0 & 45.3 & 44.6 & 39.5 & 54.0 & 50.9 & 45.0 &  0.0 &  0.0 &  0.0 &  0.7 &  4.5 & 
15.5\\
2 & 0 & 45.1 & 44.5 & 39.4 & 54.2 & 51.1 & 45.1 &  0.0 &  0.0 &  0.0 &  0.8 &  4.5 & 
15.5\\
5 &0  & 44.5 & 44.0 & 39.0 & 54.7 & 51.5 & 45.4 &  0.0 &  0.0 &  0.0 &  0.8 &  4.5 & 
15.6\\
\bottomrule
\end{tabular}

\footnotemark[2]{Others corresponds to sum of contributions from dissociative 
recombination of HCO$^+$ and CO$_2^+$ ions and resonance fluorescence of CO.}
\end{sidewaystable}

\begin{sidewaystable}[h]
\caption{The calculated contribution of different production processes 
producing \ols\ and \old\ in comet 67P for different \cod\ and CO volume mixing ratios relative to water.
\label{tab:per-o1sd}}
\scalebox{0.85}[1]{
\begin{tabular}{cclllllllllllllll}
\toprule
\multicolumn{2}{p{4cm}}{\centering Volume mixing ratios relative to water (\%)} & 
 \multicolumn{9}{c}{\centering Percentage contribution 
at different projected distances  (km)} &
\\[-22pt]
 &  &  \multicolumn{3}{c}{h$\nu$ + \wat} &  \multicolumn{3}{c}{h$\nu$  + \cod [
 h$\nu$  +OH]} 
&  \multicolumn{3}{c}{Others\footnotemark[3]} & \multicolumn{3}{c}{\grat}  \\
\cmidrule(lr){1-2} \cmidrule(lr){3-11} \cmidrule(lr){12-14}
\cod & CO & 10 & 10$^2$ & 10$^3$ & 10 & 10$^2$ & 10$^3$& 10 & 10$^2$ & 
10$^3$ 
& 10 & 10$^2$ & 10$^3$ \\ 
\cmidrule(lr){3-5} \cmidrule(lr){6-8} \cmidrule(lr){9-11} \cmidrule(lr){12-14}
1 &1  & 75.1[94.6]\footnotemark[2] & 67.3[83.9] & 44.3[54.2] & 16.4[0.8] & 14.5[5.4] & 
9.8[24.5] &  8.5[4.6] & 18.2[10.6] & 45.9[21.3] & 0.41 & 0.09 & 0.05 \\ 
1 &2 & 74.3[94.5] & 66.6[83.9] & 43.8[54.2] & 16.2[0.8] & 14.3[5.4] &  9.6[24.5] & 
9.5[4.7] & 19.1[10.7] & 46.6[21.3] & 0.42 & 0.09 & 0.05 \\
1 &5  & 71.7[94.3] & 64.2[83.6] & 42.1[54.0] & 15.6[0.8] & 13.8[5.4] &  9.3[24.5] & 
12.6[4.9] & 22.0[10.9] & 48.7[21.5] & 0.44 & 0.09 & 0.05 \\[5pt]
2 &1  & 63.3[93.5] & 57.2[82.7] & 38.7[53.5] & 27.6[0.8] & 24.6[5.4] & 17.0[24.2] &  
9.1[5.8] & 18.2[12.0] & 44.2[22.3] & 0.49 & 0.10 & 0.05 \\
2 &2  & 62.8[93.4] & 56.8[82.6] & 38.3[53.5] & 27.4[0.8] & 24.4[5.4] & 16.9[24.2] &  
9.8[5.8] & 18.9[12.0] & 44.8[22.3] & 0.49 & 0.10 & 0.05 \\
2 &5  & 61.2[93.2] & 55.2[82.4] & 37.1[53.3] & 26.6[0.8] & 23.7[5.3] & 16.3[24.1] & 
12.2[6.1] & 21.1[12.3] & 46.6[22.5] & 0.51 & 0.10 & 0.06 \\ [5pt]
5 &1  & 43.2[90.2] & 39.5[79.0] & 28.0[51.4] & 46.7[0.8] & 42.3[5.1] & 30.8[23.3] & 
10.2[9.1] & 18.2[15.8] & 41.2[25.3] & 0.69 & 0.14 & 0.07 \\ 
5 &2  & 43.0[90.1] & 39.3[79.0] & 27.9[51.4] & 46.5[0.8] & 42.2[5.1] & 30.7[23.3] & 
10.4[9.1] & 18.5[15.9] & 41.5[25.3] & 0.69 & 0.14 & 0.07 \\ 
5 & 5 & 42.5[89.9] & 38.7[78.8] & 27.3[51.3] & 45.8[0.8] & 41.6[5.1] & 30.1[23.2] & 
11.7[9.3] & 19.7[16.0] & 42.6[25.5] & 0.71 & 0.14 & 0.07 \\[5pt]
0 &1  & 92.4[95.8] & 81.7[85.3] & 51.8[54.9] &  0.0[0.8] &  0.0[5.5] &  
0.0[24.9] &  
7.6[3.4] & 18.3[9.2] & 48.2[20.2] & 0.34 & 0.07 & 0.04 \\
0 &2  &  90.9[95.7] & 80.4[85.2] & 51.0[54.9] &  0.0[0.8] &  0.0[5.5] &  0.0[24.9] &  
9.1[3.5] & 19.6[9.3] & 49.0[20.2] & 0.35 & 0.07 & 0.04 \\
0 &5  & 86.8[95.4] & 76.8[84.9] & 48.6[54.8] &  0.0[0.8] &  0.0[5.5] &  0.0[24.8] & 
13.2[3.8] & 23.2[9.6] & 51.4[20.4] & 0.36 & 0.08 & 0.04 \\ [5pt]
1 & 0 & 75.9[94.7] & 68.0[84.0] & 44.8[54.2] & 16.6[0.8] & 14.6[5.4] &  9.9[24.5] &  
7.5[4.5] & 17.4[10.6] & 45.3[21.3] & 0.41 & 0.08 & 0.05 \\
2 & 0 & 63.8[93.5] & 57.6[82.7] & 39.0[53.5] & 27.8[0.8] & 24.8[5.4] & 17.2[24.2] &  
8.4[5.7] & 17.6[11.9] & 43.8[22.4] & 0.48 & 0.10 & 0.05 \\
5 &0  & 43.2[90.2] & 39.5[79.1] & 28.1[51.4] & 46.7[0.8] & 42.4[5.1] & 30.9[23.3] & 
10.1[9.0] & 18.1[15.8] & 41.0[25.4] & 0.69 & 0.14 & 0.07 \\
\bottomrule
\end{tabular}}
\footnotemark[2]{The values in the square brackets are for \old;}
\footnotemark[3]{Others corresponds to the sum of the contributions from all reactions listed 
in  Tables 1 and 2 of \cite{Bhardwaj12} for the formation of \ols\ and \old\ 
except photodissociation of \wat\ and \cod.}
\end{sidewaystable}

\end{document}